\documentclass[twocolumn]{aastex701}

\usepackage{gensymb}
\usepackage{bm}
\usepackage{tabularx}
\usepackage{wasysym}
\usepackage[dvipsnames]{xcolor}
\shorttitle{Heterogeneities in Polarized Light}
\shortauthors{Goodis Gordon et al. 2025}
\revised{\today}
\graphicspath{{./}{figures/}}

\begin{document}

\title{Polarized Signatures of Variable Worlds: \\ Modeling Heterogeneous Habitable Earth- and Early Mars-like (Exo)planets}

\correspondingauthor{Kenneth E. Goodis Gordon}
\email[show]{Kenneth.Gordon@ucf.edu}

\author[0000-0002-4258-6703]{Kenneth E. Goodis Gordon}
\affiliation{Planetary Sciences Group, Department of Physics, University of Central Florida, 4111 Libra Drive, Orlando, FL 32816, USA}
\email{Kenneth.Gordon@ucf.edu}

\author[0000-0001-7356-6652]{Theodora Karalidi}
\affiliation{Planetary Sciences Group, Department of Physics, University of Central Florida, 4111 Libra Drive, Orlando, FL 32816, USA}
\email{tkaralidi@ucf.edu}

\author[0000-0002-4420-0560]{Kimberly M. Bott}
\affiliation{SETI Institute, Mountain View, CA 94043, USA}
\affiliation{NASA Nexus for Exoplanet System Science, Virtual Planetary Laboratory Team, Box 351580, University of Washington, Seattle, WA 98195, USA}
\email{kim.m.bott@gmail.com}

\author[0009-0007-6432-0328]{Connor J. Vancil}
\affiliation{Department of Physics, University of California, Santa Barbara, CA 93106, USA}
\email{cvancil@ucsb.edu}

\author[0000-0001-6205-9233]{Maxwell A. Millar-Blanchaer}
\affiliation{Department of Physics, University of California, Santa Barbara, CA 93106, USA}
\email{maxmb@ucsb.edu}

\author[0000-0002-0413-3308]{Nicholas F. Wogan}
\affiliation{Space Science Division, NASA Ames Research Center, Moffett Field, CA 94035, USA}
\affiliation{NASA Nexus for Exoplanet System Science, Virtual Planetary Laboratory Team, Box 351580, University of Washington, Seattle, WA 98195, USA}
\email{nicholaswogan@gmail.com}

\author[0000-0002-7188-1648]{Eric T. Wolf}
\affiliation{Laboratory for Atmospheric and Space Physics, University of Colorado Boulder, Boulder, CO 80303, USA}
\affiliation{Blue Marble Space Institute of Science, Seattle, WA 98104, USA}
\affiliation{Consortium on Habitability and Atmospheres of M-dwarf Planets (CHAMPs), Laurel, MD 20723, USA}
\affiliation{NASA Nexus for Exoplanet System Science, Virtual Planetary Laboratory Team, Box 351580, University of Washington, Seattle, WA 98195, USA}
\email{eric.wolf@colorado.edu}


\begin{abstract}

Determining the habitability of terrestrial exoplanets is a complex problem that represents the next major step for the astrophysical community. The majority of current models treat these planets as homogeneous or contain heterogeneity that is constant in time. In reality, habitable exoplanets are expected to contain atmospheric and surface heterogeneities similar to Earth, with diurnal rotation, seasonal changes, and weather patterns resulting in complex, time-dependent signatures. Due to its ability to measure light as a vector, polarimetry provides an important tool that will enhance the characterizations of heterogeneous worlds. Here we model the visible to near-infrared linear spectropolarimetric signatures, as functions of wavelength and planetary phase angle, of various heterogeneous Earth scenarios as well as the first signals of an early wet and potentially habitable Mars. The contributions from the different atmospheric and surface properties result in asymmetric phase curves and variable spectra, with the polarization appearing to be more sensitive than flux to heterogeneities such as patchy clouds and continents moving into and out-of-view. Our models provide important predictions of expected polarized and unpolarized signatures of heterogeneous exoplanets that will help guide the designs and observing plans of future polarimeters, including those proposed for the upcoming Habitable Worlds Observatory.

\end{abstract}



\section{Introduction} \label{sec: Intro}

A commonality of planets across the solar system is that they possess time-variable heterogeneous features in their atmospheres and surfaces. Observations have shown, for example, variability over time in the bands of Jupiter \citep[e.g.,][]{ingersoll2004}, the clouds and surface of Titan \citep[e.g.,][]{mitchell2016}, the planetary-wide dust storms \citep[e.g.,][]{cantor2001} and polar ice caps \citep[e.g.,][]{byrne2009} of Mars, and weather patterns of the Earth \citep[e.g.,][]{ruddiman2001}. Such observations reveal that planetary heterogeneity leaves clear traces on the resulting disk-integrated signals of these objects when observed across multiple phases \citep[i.e., time-resolved observations; e.g.,][]{cowan2011, ge2019}.

These disk-integrated signals can vary due to the diurnal rotation of the planet \citep[e.g.,][]{cowan2009, robinson2011, robinson2018, lustigyaeger2018}, atmospheric dynamics \citep[i.e., weather; e.g.,][]{roccetti2025}, and the seasonal changes throughout the planet's orbit around its parent star. For example, \citet[][]{livengood2011} used observations of the Earth-as-an-exoplanet from the \textit{EPOXI} spacecraft to constrain the heterogeneity of the Earth's surface. Their analyses showed clear diurnal variability in the resulting planetary flux over the course of one Earth day, while seasonal differences were also detectable between the different observing epochs taken from multiple points in Earth's orbit.

Given how prevalent heterogeneity is throughout the solar system, it is expected that exoplanets will exhibit similar trends. However, for computational efficiency, most models of exoplanets treat planets as homogeneous objects \citep[e.g.,][]{stam2008, karalidi2011}, or if they include heterogeneity, they do so only for a single or a few select planetary configurations, and thus the signals are still constant in time \citep[i.e., "frozen" heterogeneity;][]{sanroma2013, trees2022, gomezbarrientos2023, gordon2023}. Ignoring the effects of variability will lead to errors in the characterization of these planets \citep[e.g.,][]{kofman2024, kelkar2025, roccetti2025}.

Constraining the heterogeneity of exoplanets is especially crucial for habitability studies, as the search for life outside the solar system depends on comparisons of the exoplanet signals to those of Earth. However, the Earth environment is a dynamic rather than a static system, and the heterogeneity of the atmosphere and surface of the Earth is considered key to the emergence and existence of life on this planet \citep[e.g.,][]{foley2015, glaser2020}. Although the habitability of an exoplanet is largely tied to the existence of liquid water on its surface \citep[e.g.,][]{kopparapu2013, kopparapu2016, wolf2017}, life also needs land to develop and thrive on \citep[e.g.,][]{lingam2019} and so a ``balanced" blend of ocean and land is now believed to be necessary for a planet to be habitable and to be detected as such.

The majority of current exoplanet characterizations rely solely on the reflected flux (i.e., unpolarized light) from these planets. However, flux-only observations use only a fraction of the informational content of the light (i.e., the Stokes I), resulting in degeneracies in the characterizations of the planetary atmospheres and surfaces \citep[e.g.,][]{hansenhovenier1974, stam2005, karalidi2011, karalidi2012rainbow, west2022}. On the other hand, polarization takes into account the full informational content of the light (i.e., the full Stokes vector) and is thus sensitive to the micro- and macro-physical properties of the planet that cause the scattering of the light \citep[e.g.,][]{hansenhovenier1974, west1991}. Spectropolarimetry and polarimetric photometry can therefore complement flux observations to break degeneracies and better constrain planetary atmospheres and surfaces \citep[e.g.,][]{mishchenko1994, stam2005, miles2014, emde2018, sterzik2020, gordon2023}.

Due to its vector nature, polarimetry is also sensitive to the locations of specific features on the observed planetary disk. Whereas flux-only observations introduce a number of degeneracies in the characterizations of planetary features, such as being unable to distinguish between a small surface feature at the center of the planet versus a larger feature at higher latitudes \citep[e.g.,][]{karalidistam2012}, polarimetry would allow for the correct characterization of the planetary heterogeneity by being able to separate these two cases. Polarization is thus crucial for mapping heterogeneous exoplanets and is vital for disentangling various habitability indicators such as forward scattering from water clouds versus ocean glint \citep[e.g.,][]{robinson2011, lustigyaeger2018}, the latitude albedo effect of ocean glint \citep[e.g.,][]{cowan2012}, and evidence of continents \citep[e.g.,][]{foley2015}.

While several studies have analyzed the effects of atmospheric asymmetries on the polarized signals of gas giant exoplanets and brown dwarfs \citep[e.g.,][]{sengupta2010, dekok2011, karalidi2013flux, stolker2017, chubb2024}, fewer studies have investigated the effects of heterogeneous features on the resulting polarimetric signals of terrestrial exoplanets. \citet[][]{karalidistam2012} and \citet{karalidi2012rainbow}, e.g., modeled horizontally and vertically heterogeneous planets with different surface and cloud maps, including water ice clouds, and showed the importance of polarization in characterizing these planets and determining the locations of features across the disk. However, these models did not account for any temporal variability (i.e., they modeled ``frozen heterogeneous" planets) and only calculated the broadband flux and polarimetric signals of the planets. \citet{treesstam2019} and \citet{trees2022} expanded upon these earlier studies to analyze the effects of realistic wavy oceans and ``dry land" of different albedos on the flux and polarimetric signals (both broadband and spectra) of Earth-like exoplanets.
Recently, \citet[][]{roccetti2025} used the realistic wavelength-dependent surface albedo map HAMSTER \citep[][]{roccetti2024}, as well as heterogeneous 3D cloud maps with sub-grid cloud variability, to study the flux and polarimetric spectra of the Earth-as-an-exoplanet. This study highlighted the importance of accounting for seasonal variability, as well as the impact of the model grid size, on the planetary signals. However, \citet[][]{roccetti2025} did not address the impact of diurnal rotation on the model signals.

Additionally, most previous studies for terrestrial exoplanet polarization focused only on the signals of modern Earth, or modern Earth-like planets. \citet{goodisgordon2025}, however, showed that the spectropolarimetric signal of Earth has changed considerably throughout its life, as the planet evolved through periods of habitability and non-habitability. Being able to differentiate the signals of an Earth-like planet through its evolution, and to distinguish it from other terrestrial (and potentially habitable) planets such as an early (wet) or modern (dusty) Mars in both unpolarized and polarized light, will be crucial for ground- and space-based missions targeted at characterizing terrestrial exoplanets, including the upcoming Extremely Large Telescopes (ELTs) and the Habitable Worlds Observatory (HWO).

Here, we build upon these previous studies by investigating the effects of time-variable heterogeneous features, including diurnal rotation and seasonal variations, on the resulting signals of Earth and Mars at different times in their evolution. We used an advanced polarization-enabled radiative transfer code to model the unpolarized and polarized visible to near-infrared (VNIR) reflected flux of the planets as functions of both wavelength $\lambda$ and planetary phase angle $\alpha$. All of our models assume circular edge-on orbits (i.e., eccentricity = 0 and inclination = 90$\degree$) with the planets orbiting 1 au from the Sun. Our models are publicly available on Zenodo (\dataset[doi:10.5281/zenodo.17039224]{\doi{10.5281/zenodo.17039224}}) for the community to assist in characterizing terrestrial exoplanets.

This paper is organized as follows. In Sect.~\ref{sec: RefPol} we provide definitions of reflected light polarization and explain the numerical method used to calculate the signals presented here. In Sect.~\ref{sec: Models} we discuss the atmospheric, surface, and aerosol properties we used in our models and provide justifications for these various inputs. In Sect.~\ref{sec: Hetero}, we present our model spectra and phase curves in both unpolarized and polarized light and discuss the effects of various types of planetary heterogeneities. In Sect.~\ref{sec: Observe} we provide first-order observing constraints, based on our models, for upcoming spectropolarimeters aimed at characterizing terrestrial exoplanets. Finally, in Sect.~\ref{sec: DiscussConclude} we discuss and summarize our results and present a future outlook.

\section{Reflected Light Polarimetry} \label{sec: RefPol}

\subsection{Calculating Flux and Polarization} \label{sec: defs}

Reflected light from a planet can be fully described by a flux vector $\pi\textbf{F}$ \citep[see, e.g.,][]{hansentravis1974, hovenier2004, stam2008}

\begin{eqnarray}
\pi\textbf{F}  =  \pi\left[\begin{array}{c} F \\
Q \\
U \\
V\end{array}\right],
\label{eq:fluxvec}
\end{eqnarray}
with $\pi$F the total, $\pi$Q and $\pi$U the linearly polarized, and $\pi$V the circularly polarized fluxes. All four parameters are wavelength-dependent and have units of $W$ $m^{-2} m^{-1}$. Fluxes $\pi$Q and $\pi$U are defined with respect to the planetary scattering plane \citep[i.e., the plane through the centers of the host star, planet, and observer; see, e.g.,][]{stam2008, gordon2023}. Note that this plane is not necessarily the same as the orbital plane for the planet. The two planes would only coincide at all phase angles for edge-on (i.e., i = 90$\degree$) planetary orbits. Here, we ignore $\pi$V, since studies have shown that the circular polarization from exoplanets will be negligible \citep[e.g.,][]{hansentravis1974, rossi2018}.

The degree of linear polarization, $P$, of flux vector $\pi\textbf{F}$ is defined as

\begin{eqnarray}
P = \frac{\sqrt{Q^{2} + U^{2}}}{F}
\label{eq:degofpol}
\end{eqnarray}
and is independent of the reference plane.

\subsection{Calculating the Heterogeneous Signals} \label{sec: code}

We used the Doubling Adding Program (hereafter DAP) polarization-enabled radiative transfer code to calculate the signatures of our model planets. Our code fully incorporates single and multiple scattering by atmospheric gases as well as aerosol and cloud particles and can model atmospheres of any composition with as many layers as needed. For more information on our code, we refer the reader to \citet[][]{gordon2023, goodisgordon2025} and references therein.

DAP defines the flux vector of stellar light that has been reflected by a spherical planet with radius $r$ at a distance $d$ from the observer (where $d \gg r$) as

\begin{eqnarray}
\pi\textbf{F}(\lambda, \alpha)  =  \frac{1}{4}\frac{r^2}{d^2}\textbf{S}(\lambda, \alpha)\pi\textbf{F}_0(\lambda),
\label{eq:fluxplanet}
\end{eqnarray}
where $\lambda$ is the wavelength of the light and $\alpha$ is the planetary phase angle. $\pi\textbf{F}_0$ is the flux vector of the incident starlight and $\textbf{S}$ is the $4 \times 4$ planetary scattering matrix. For our calculations, we normalized Eq.~\ref{eq:fluxplanet} assuming $r = 1$ and $d = 1$.

Numerous past studies that used DAP simulated horizontally homogeneous planets, or if they simulated heterogeneity, they did so using the weighted sum approximation \citep[see, e.g.,][]{stam2008, gordon2023}. Here, we are interested in studying the effects of time variability on realistic horizontally and vertically heterogeneous planets. Therefore, we used the version of DAP described in \citet[][]{karalidistam2012} that was modified to model spatially resolved, horizontally and vertically heterogeneous planets. The planet is divided into pixels that are small enough for the local atmosphere and surface to be considered both plane-parallel and horizontally homogeneous. For each unique pixel, the code first calculates the local reflection matrix $\textbf{R}$ and then, taking into account the location of each pixel in the star-planet-observer system, it sums the signal of all pixels to calculate \textbf{S} \citep[see][for more details]{karalidistam2012}:

\begin{eqnarray}
\textbf{S}(\alpha)  =  \frac{4}{\pi}\sum_{i=1}^{N} \mu_{i}\mu_{0i}\textbf{L}(\beta_{2i}) \textbf{R}_{i}(\mu_{i},\mu_{0i},\Delta\phi_{i}) \textbf{L}(\beta_{1i}) dO_{i},
\label{eq:scattermatr}
\end{eqnarray}
where $N$ is the total number of pixels on the illuminated part of the planetary disk, $dO$ is a surface element on the planet, and $\textbf{L}$ are the rotation matrices used to rotate from the planetary scattering plane to the local meridian planes and back. For more information on the process and geometry, as well as the validation of the code, we refer the reader to \citet[][]{karalidistam2012}.

Here we assumed unpolarized incident starlight, based on observations of the Sun \citep[e.g.,][]{kemp1987} and other active and inactive FGK stars \citep[e.g.,][]{cotton2017}. In this case, $\pi\textbf{F}_0$ = $\pi$$F_0$\textbf{1}, where \textbf{1} is the unit column vector and $\pi$$F_0$ is the total incident stellar flux (measured perpendicular to the direction of propagation of the light), for which we assumed a normalized value of 1 $W$ $m^{-2} m^{-1}$ \citep[see, e.g.,][]{karalidistam2012}. Due to these normalizations and assumption of unpolarized incident starlight, rotation matrix $\textbf{L}(\beta_{1i})$ can be ignored in Eq.~\ref{eq:scattermatr} and the total reflected flux (cf. Eq.~\ref{eq:fluxplanet}) becomes

\begin{eqnarray}
\pi{F_{n}}(\lambda, \alpha)  =  \frac{1}{4}a_{11}(\lambda, \alpha)
\label{eq:normalizedflux}
\end{eqnarray}
where $a_{11}$ is the (1,1)-element of $\textbf{S}$ and the subscript $n$ on the flux indicates the normalization \citep[see, e.g.,][]{stam2008, karalidi2011}.

\section{Model Descriptions} \label{sec: Models}

To capture the horizontal heterogeneity of our planets, we ran all simulations with a spatial resolution of $2\degree \times 2\degree$, resulting in 8100 pixels across the full observable planetary disk. Using our heterogeneous radiative transfer code (see Sect.~\ref{sec: code}), the full spectropolarimetric signals of the planets were generated for wavelengths ($\lambda$) between 0.4 and 1.8~$\mu$m, covering the expected range for the upcoming HWO \citep[e.g.,][]{mamajek2024}. For computational efficiency, we limit our phase angle ($\alpha$) coverage to 18 distinct phases between $30\degree$ and 120$\degree$, which covers phases over which we expect to observe key habitability features of Earth-like planets with upcoming observatories \citep[e.g.,][]{vaughan2023}.

\subsection{Model Surfaces} \label{sec: Surfs}

Our surfaces are modeled as Lambertian (i.e., isotropically reflecting and depolarizing) surfaces with wavelength-dependent (unless otherwise stated) albedos, as is commonly done for Earth-like modeling and retrievals \citep[see, e.g.,][and references therein]{tilstra2021}. We modeled six different categories of surfaces: ocean, forest (a combination of deciduous and conifer), grass, sand, snow/ice, and fresh basalt. The reflection properties of these surfaces were taken from the NASA JPL EcoStress Spectral Library\footnote{\url{https://speclib.jpl.nasa.gov}} \citep[][]{baldridge2009, meerdink2019} as well as the USGS Spectral Library\footnote{\url{https://crustal.usgs.gov/speclab/QueryAll07a.php}} \citep[][]{kokaly2017}. For our models of an early, potentially habitable Mars (see Sect.~\ref{sec: EarlyMars}) we  used a wavelength-independent broadband albedo of 0.15 for the bare soil of this planet, following \citet[][]{guzewich2021}.

\subsection{Model Atmospheres} \label{sec: Atmos}

For our atmospheric profiles, we used inputs from multiple sources, including both 1D and 3D simulations, to model the modern and early atmospheres of Earth and Mars. Justifications for the atmospheric makeups, including their gaseous vertical mixing ratios (VMRs) and aerosol properties, are described in the following subsections. With the exception of our early Mars models (see Sect.~\ref{sec: EarlyMars}), all pixels in each of our simulated planets used a single appropriate T-P profile and set of VMRs for that planetary scenario. All of our sources used the wavelength-dependent solar evolution correction of \citet[][]{claire2012} as the input stellar source for their atmospheric calculations.

\subsection{Our Planetary Scenarios}

\subsubsection{Modern Earth} \label{sec: ModernEarth}

Our modern Earth model was adopted from \citet[][]{goodisgordon2025} and simulates the Earth as it appears today, with an N$_2$-dominated atmosphere containing 21\% O$_2$ and $\sim$0.0366\% CO$_2$, followed by present-day trace amounts of CO, CH$_4$, O$_3$, N$_2$O, and NO. It was generated using the 1D photochemical-climate code \textit{Atmos} \citep[see, e.g.,][and references therein]{arney2016, arney2017, arney2018}, following the chemical network described in \citet[][]{arney2016}. As explained in \citet[][]{goodisgordon2025}, we binned the original 200-layer atmosphere down to 45 layers for computational efficiency.

Our heterogeneous modern Earth model contains pixels with both clear (i.e., cloud-free) and cloudy atmospheres. Our cloudy models contained either liquid water stratocumulus clouds or ice water cirrus clouds. The clouds were placed in the appropriate atmospheric layer corresponding to altitudes of $\sim$1 km and $\sim$10 km, respectively. For simplicity, the optical thicknesses ($\tau$) of our stratocumulus clouds were set to 10 and those of our cirrus clouds were set to 0.5 for all models, based on the average properties derived from MODIS (Moderate Resolution Imaging Spectroradiometer) observations \citep[e.g.,][]{king2004, king2013}. We modeled the optical properties of these clouds using Mie theory \citep[][]{derooijvanderstap1984}, which provides a good first-order approximation and improves computational runtime \citep[e.g.,][]{schmid2011, mclean2017}. Our cloud particles used the standard two-parameter gamma particle size distribution of \citet[][]{hansenhovenier1974}. The liquid water droplets had an effective radius r$_{eff}$ = 6 $\mu$m and effective variance u$_{eff}$ = 0.4, while the water ice particles had r$_{eff}$ = 10 $\mu$m and u$_{eff}$ = 0.1 \citep[see, e.g.,][]{vandiedenhoven2007, goodisgordon2025}. We used the wavelength-dependent complex refractive indices from \citet[][]{hale1973} and \citet[][]{warren2008} for the liquid and ice water clouds, respectively. We acknowledge that water ice crystals are nonspherical in nature and can produce different optical properties than spherical Mie-scattering particles \citep[e.g.,][]{heymsfield1984, bar1988}. For example, using spherical cloud particles can introduce extraneous features in the planetary phase curves, such as primary rainbows, that would not otherwise be present for planets with atmospheres containing only nonspherical cloud particles. We refer the reader to \citet[][]{goodisgordon2025} and the references therein for more information on the impacts of our choice to adopt Mie vs. nonspherical scattering particles.

\subsubsection{Snowball Earth} \label{sec: SnowballEarth}

To simulate another habitable point of Earth's history that is drastically different from today, we generated models for a so-called ``Snowball Earth" event \citep[see, e.g.,][for descriptions of the Snowball Earth hypothesis]{kirschvink1992, hoffman1998}. In particular, we simulated how Earth would have appeared $\sim$700 million years ago during the Sturtian Glaciation event \citep[e.g.,][]{hoffman2017}. The atmospheric properties of these models were calculated using the 1D photochemical-climate code \textit{Photochem}, following the chemical network of \citet[][]{wogan2023} in which iterations occur between photochemistry and climate models until a full steady state is achieved. The code assumed an initial surface albedo of 0.65, following the ice-albedo feedback parameterization in \citet[][]{arney2016} that captures the reflectiveness of both the surface ice and any possible water clouds. Upon reaching steady state, the surface temperature of the model ended up being 227 K, which is a reasonable global average temperature for a Snowball Earth event \citep[e.g.,][]{abbot2011}.

The atmosphere is N$_2$-dominated with $\sim$0.1\% CO$_2$ and $\sim$0.2\% O$_2$, following Figure 5 of \citet[][]{catling2020} for 700 million years ago. H$_2$O evaporates from the surface with a relative humidity in the atmosphere of 100\%. A surface CH$_4$ flux into the atmosphere of $\sim$30 Tmol/yr is imposed, which is the same as on modern Earth. Several studies have contested the necessity of water clouds to deglaciate a Snowball Earth \citep[e.g.,][]{abbot2012, abbot2014, braun2022}. Therefore, in addition to cloud-free atmosphere simulations, we also simulated Snowball Earth atmospheres containing water clouds. Due to the low atmospheric temperatures of our Snowball Earth scenario, the majority of the clouds for these models were low-altitude, optically thick \citep[$\tau$ = 10; see, e.g.,][]{abbot2014} water ice clouds containing similar particles to our modern Earth models (see Sect.~\ref{sec: ModernEarth}). In Sect.~\ref{sec: Weather} we discuss in more detail the effects that different distributions of water clouds have on our resulting heterogeneous Snowball Earth signals.

\subsubsection{Early Mars} \label{sec: EarlyMars}

While Earth is the only planet known to be currently habitable, abundant evidence suggests that liquid water was widespread on ancient Mars during its Noachian period ($\sim$4.1 - 3.7 Ga), thus leading to habitable conditions \citep[see, e.g.,][for an overview]{wordsworth2016}. For our early Mars model we used the ``RW2TPWlakes100" simulation of \citet[][]{guzewich2021}, which was generated using the ROCKE-3D GCM code \citep[][]{way2017} and simulated Mars as it appeared 3.8 Ga. The surface used the inferred paleotopography map of \citet[][]{bouley2016} and was initialized with a 100 m global equivalent layer of existing surface liquid water as ``lakes" in topographic low points.

The atmosphere was composed of 40 vertical layers and followed \citet[][]{wordsworth2017} for the atmospheric VMRs ($\sim$94\% CO$_2$, $\sim$5\% H$_2$, and $\sim$1\% CH$_4$ as well as trace amounts of H$_2$O) and temperature-pressure profile. Dust was not included in the simulated atmosphere, as Martian dust probably formed after any wet climate period \citep[see][and references therein, for more details]{guzewich2021}. Although some studies have suggested that CO$_2$ ice clouds may have assisted in warming the surface of ancient Mars \citep[e.g.,][]{wordsworth2013}, at the time the models of \citet[][]{guzewich2021} were created, CO$_2$ cloud physics were still in development for ROCKE-3D and so only H$_2$O clouds were included in the simulation. Similarly to our two Earth scenarios (see Sect.~\ref{sec: ModernEarth} and \ref{sec: SnowballEarth}), we modeled these water ice particles as Mie scatterers. However, unlike the Earth scenario clouds, the ice cloud content per atmospheric layer for each pixel in our early Mars model was allowed to vary following the ROCKE-3D model. For simplicity and computational efficiency, we averaged the vertical ice cloud contents into three distinct latitudinal regions across the planet: a North Polar (NP) region extending from 90$\degree$ N to 50$\degree$ N, a South Polar (SP) region extending from 90$\degree$ S to 50$\degree$ S, and an Equatorial (EQ) region covering 50$\degree$ N to 50$\degree$ S (see right panel of Fig.~\ref{fig: Marsviews}). The ice cloud r$_{eff}$ were unrestricted by the published ROCKE-3D model, as they were not part of the standard output at the time. We therefore opted to use the r$_{eff}$ of modern Mars ice clouds \citep{guzewich2014,guzewich2019}, as these values provided the best published global distributions of ice cloud particle sizes for Mars. In particular, we used an altitude-constant r$_{eff}$ = 1.5 $\mu$m for the polar regions and an altitude-dependent r$_{eff}$ ranging from 1.5 - 3.0 $\mu$m for the equatorial region. We acknowledge that since the early Mars atmosphere was warmer and denser than the planet's modern-day atmosphere, early Mars water ice clouds could have contained larger particles, which could alter the features of the clouds in the planetary signals. However, as no predictions exist today for the global distributions of cloud particle sizes on early Mars, we used modern Mars values to reduce the number of assumptions we made in our models.

\section{Effects of Different Heterogeneities} \label{sec: Hetero}

In this section, we investigate the effects that different time-variable planetary phenomena have on the disk-integrated normalized flux ($\pi{F_{n}}$) and degree of linear polarization ($P$) signatures of our models. Unless otherwise stated, our model spectra cover wavelengths from $\lambda$ = 0.4 - 1.8 $\mu$m and have a variable spectral resolving power, R $(= \lambda / \Delta\lambda)$, with R$\sim$5 from 0.4 – 0.6 $\mu$m, R$\sim$65 to $\sim$150 from 0.65 - 1.5 $\mu$m, and R$\sim$30 from 1.55 – 1.8 $\mu$m. Our model phase curves capture 18 key orbital phases between 30$\degree$ and 120$\degree$.

\subsection{Diurnal Rotation} \label{sec: Rotate}

\begin{figure}[ht!]
    \centering
    \includegraphics[width=\linewidth]{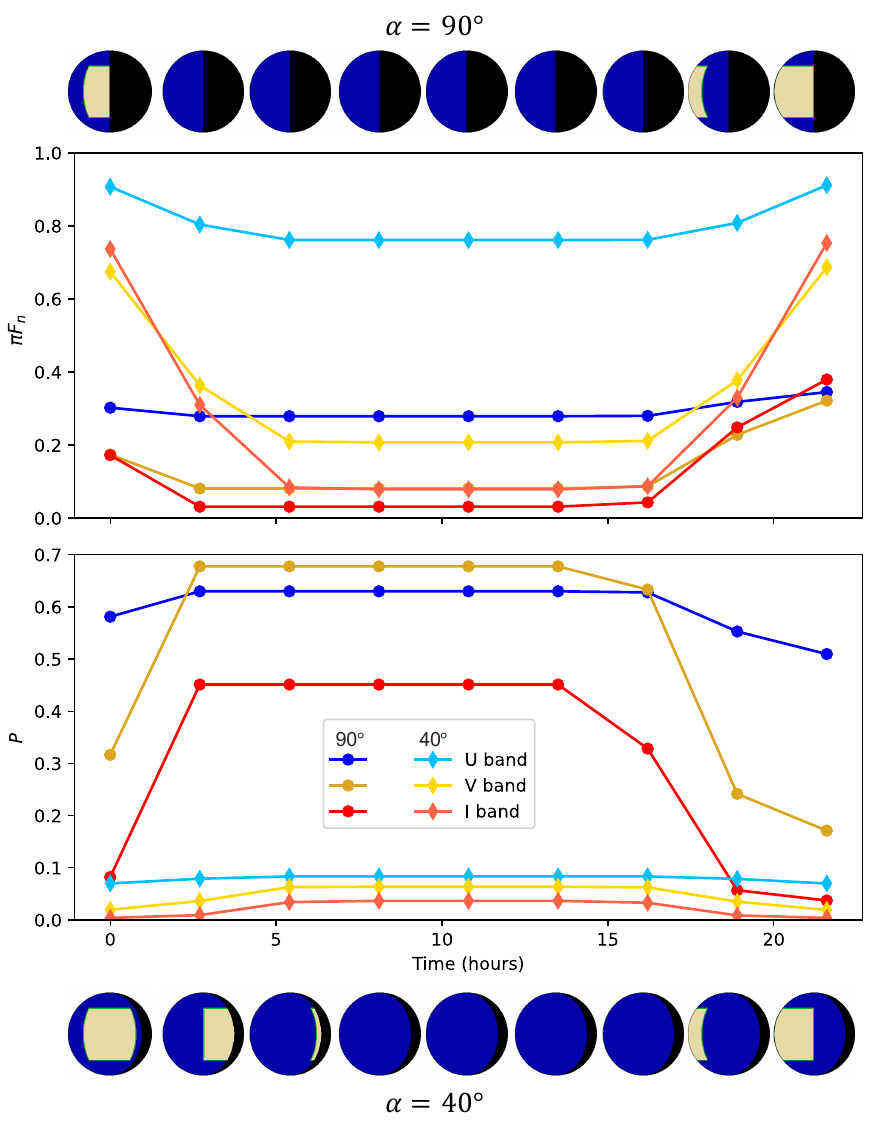}
    \caption{Diurnal rotation light curves of a planet with a Lambertian reflecting ocean, a single large sandy supercontinent, and a clear Proterozoic Earth atmosphere \citep[see][]{goodisgordon2025} observed at phase angles of $\alpha$ = 90$\degree$ (circles) or $\alpha$ = 40$\degree$ (diamonds) at $\sim$2.7 hour increments. The supercontinent rotating in and out of view leaves clear traces on both the disk-integrated reflected flux, $\pi{F_{n}}$, (top) and degree of linear polarization, $P$, (bottom) of the planet.}
    \label{fig: SCmodels}
\end{figure}

The diurnal rotation of a planet, with different surface and atmospheric features moving into and out of view, can lead to detectable variations in the resulting planetary flux. Here we analyze how these changes also affect the resulting polarization of the planets. As a simplified test case, Fig.~\ref{fig: SCmodels} shows the $\pi{F_{n}}$ (top) and $P$ (bottom) diurnal light curves for an ocean planet with a single sandy supercontinent moving into and out of view. The atmospheric inputs for this model were taken from the \citet[][]{goodisgordon2025} Proterozoic Earth scenario with clear, cloud-free atmospheres. We ``observe'' the planet at nine different points over one full 24-hour rotation in the U-band (blue points), V-band (gold points), and I-band (red points). The planet is observed at either quadrature (i.e., $\alpha = 90\degree$, circles), which provides the largest angular separation between the planet and its host star, or $\alpha = 40\degree$ (diamonds), where the liquid water rainbow feature appears \citep[e.g.,][]{bailey2007}.

As expected, the observed light curves show larger $P$ at 90$\degree$ compared to 40$\degree$ due to the Rayleigh scattering \citep[e.g.,][]{hansenhovenier1974, stam2008}. On the other hand, the planet is brighter for $\alpha = 40\degree$ due to the larger illuminated surface area. In both $\pi{F_{n}}$ and $P$ we see larger variations in the signals from shorter (blue) to longer (red) $\lambda$, as the atmospheric optical thickness changes and the light probes deeper through the atmosphere, therefore allowing us to see more effects from the rotating surface. The higher reflectivity of the sandy continent compared to the darker ocean leads to higher $\pi{F_{n}}$ and lower $P$ when the continent dominates the observable disk. When the continent moves out of view and only the ocean is visible (i.e., time steps 2 through 6 for $\alpha = 90\degree$ and 3 through 5 for $\alpha = 40\degree$), the signals remain constant. The $\pi{F_{n}}$ light curves show larger separation between the bandpasses at $\alpha = 40\degree$ than at $\alpha = 90\degree$, while the $P$ light curves show the opposite trend, thus highlighting the importance of observing both flux and polarization to better understand the variability due to the planetary rotation. We acknowledge that if we were to model our ocean surface with Fresnel reflection and rough waves \citep[see, e.g.,][]{treesstam2019, trees2022} rather than with a smooth Lambertian surface, we could see additional effects from the ocean glint in our resulting signals. However, because the disk-integrated nature of exoplanetary signals is expected to blur the surface reflectance \citep[see, e.g.,][]{kopparla2018, gordon2023, goodisgordon2025}, modeling Fresnel reflecting surfaces is outside the scope of this study.

\begin{figure}[ht!]
    \centering
    \includegraphics[width=\linewidth]{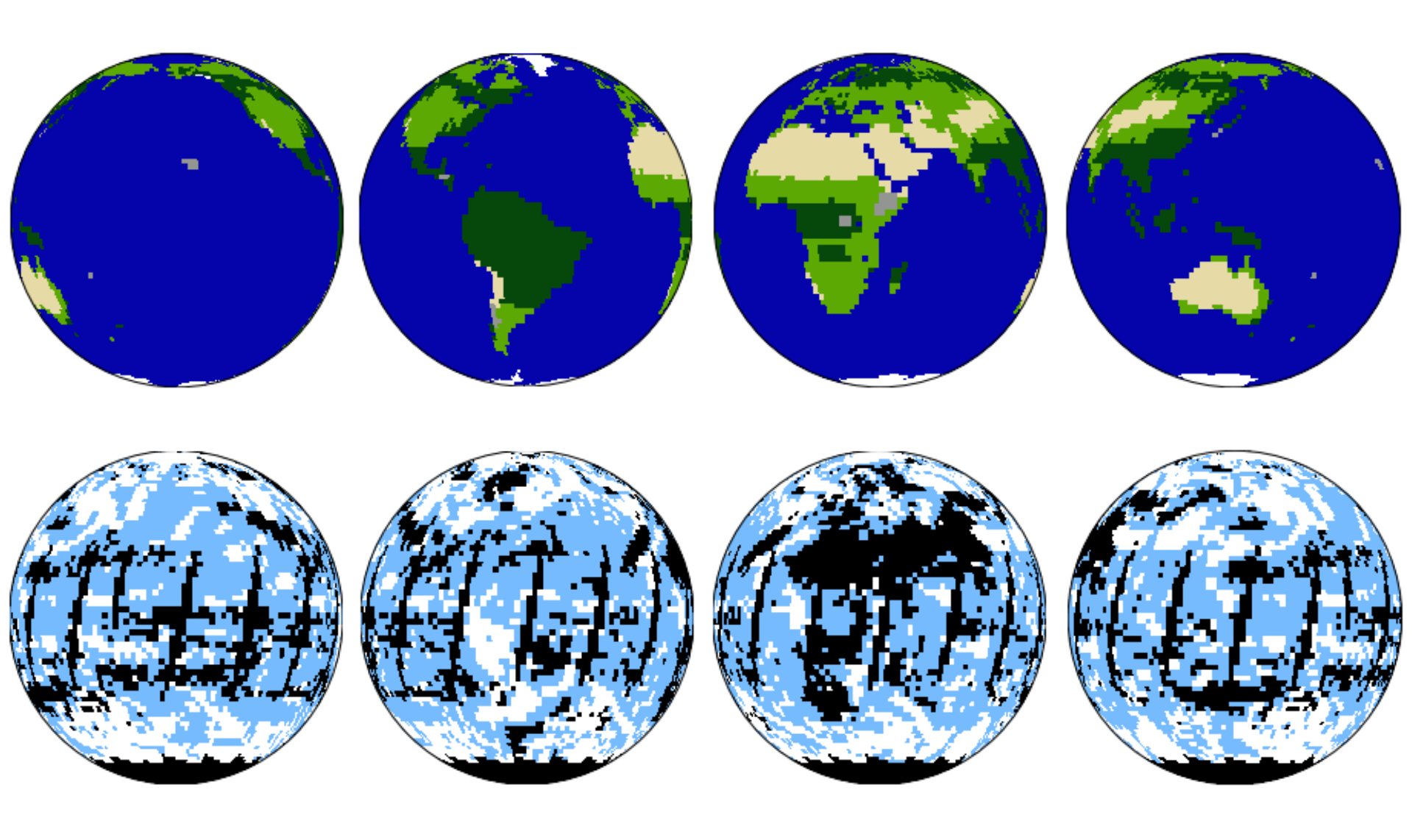}
    \caption{Our model views of modern Earth throughout its diurnal rotation, with either the Pacific Ocean (left), the Amazon rainforest (middle left), the Horn of Africa (middle right), or Papua New Guinea (right) centered on the disk. The top row shows the surface distributions while the bottom row shows the corresponding cloud distributions we used (see Sect.~\ref{sec: Atmos}), with cirrus clouds in light blue, stratocumulus clouds in white, and no clouds (i.e., clear atmosphere) in black.}
    \label{fig: ModEarthviews}
\end{figure}

We next simulated the diurnal rotation of the modern Earth. Fig.~\ref{fig: ModEarthviews} shows four views of the planet over the course of one day, with either the Pacific ocean (longitude 159$\degree$ W), Amazon rainforest (longitude 59$\degree$ W), Horn of Africa (longitude 41$\degree$ E), or Papua New Guinea (longitude 141$\degree$ E) at the center of the illuminated portion of the globe. Our models include both liquid water and water ice clouds distributed across the planet based on MODIS observations taken on 18 May 2013 (see Sect.~\ref{sec: ModernEarth} for descriptions of the cloud properties).

\begin{figure*}[ht!]
    \centering
    \includegraphics[width=11.5cm]{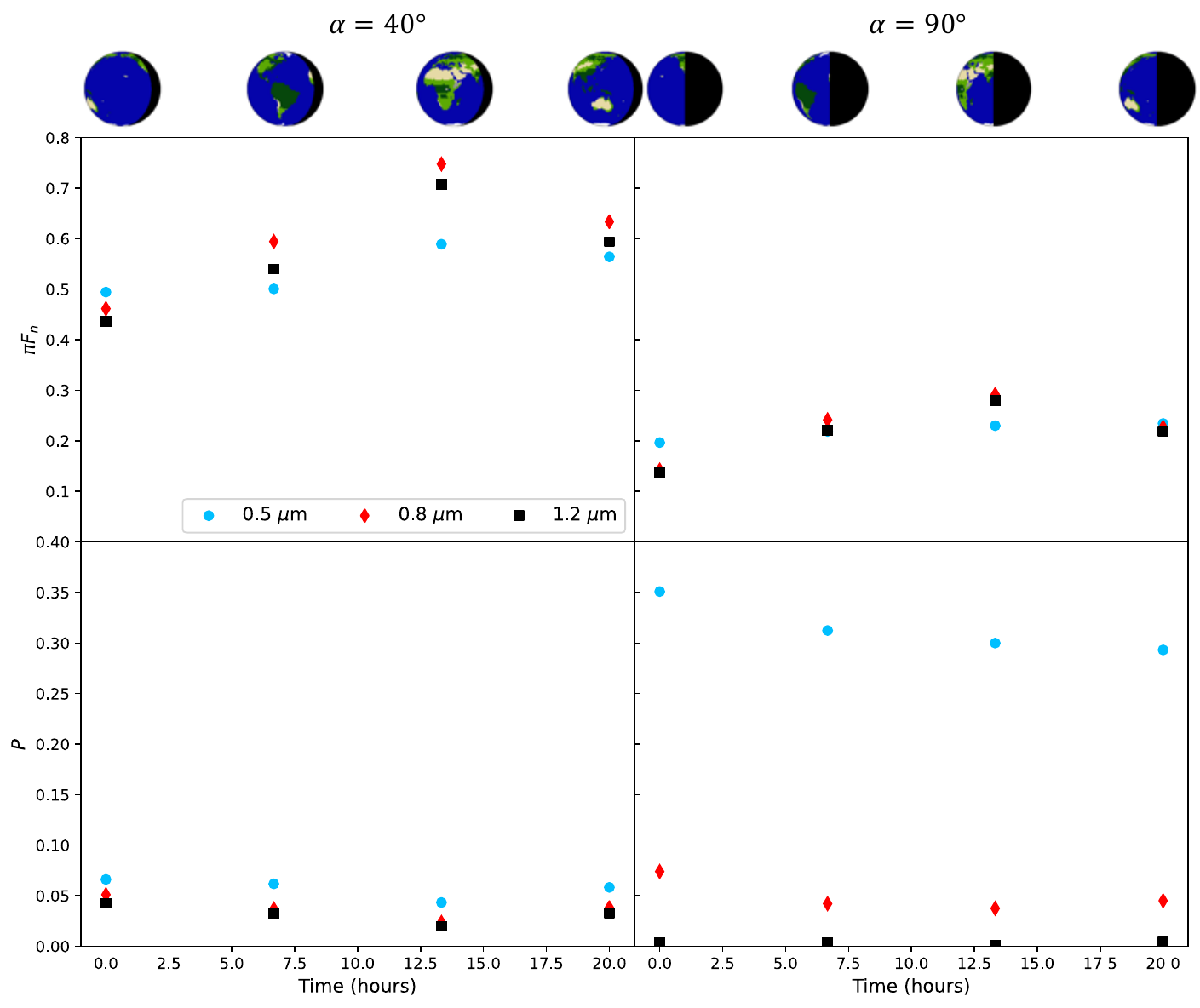}
    \caption{Diurnal rotation of our modern Earth scenario observed at three different $\lambda$ and either $\alpha = 40\degree$ (left) or $\alpha = 90\degree$ (right). Snapshots of the surface distributions for each view at both phases are shown in the top images. While the reflected total normalized flux, $\pi{F_{n}}$ (top), shows overlapping variability at both phases between the observed bandpasses as the planet rotates, we can see clear separation between observations at $\alpha = 90\degree$ in the degree of linear polarization, $P$, curves (bottom).}
    \label{fig: ModEarthrotate}
\end{figure*}

In Fig.~\ref{fig: ModEarthrotate} we display the $\pi{F_{n}}$ (top) and $P$ (bottom) signals for one rotation of the modern Earth at three different wavelengths ($\lambda = 0.5$ $\mu$m, blue circles; $\lambda = 0.8$ $\mu$m, red diamonds; $\lambda = 1.2$ $\mu$m, black squares) and for either $\alpha = 40\degree$ (left) or $\alpha = 90\degree$ (right). We can see clear variability in both $\pi{F_{n}}$ and $P$ as different surfaces dominate the field of view, with, e.g., the highly reflective sandy surfaces of the Sahara desert leading to larger $\pi{F_{n}}$ than the darker Pacific ocean. As more of the planetary surface enters the shadow at $\alpha = 90\degree$, the level of variability in the flux decreases greatly, and the signals across the different wavelength bands become overlapped, thereby making it difficult to separate and distinguish the variability. On the other hand, the polarized signals show clearer separation between the different bandpasses, especially at $\alpha = 90\degree$ due to Rayleigh scattering. Additionally, while the variability in the signals in $\pi{F_{n}}$ change across wavelength (i.e., more flux at $\lambda = 0.5$ $\mu$m for the Pacific view versus more flux at $\lambda = 0.8$ $\mu$m for the Africa view), the variability in the signals for $P$ are wavelength independent. This would imply that we would only need to observe the planet in one bandpass in polarized light to acquire an understanding of the planetary variability.

\begin{figure*}[ht!]
    \centering
    \includegraphics[width=11.5cm]{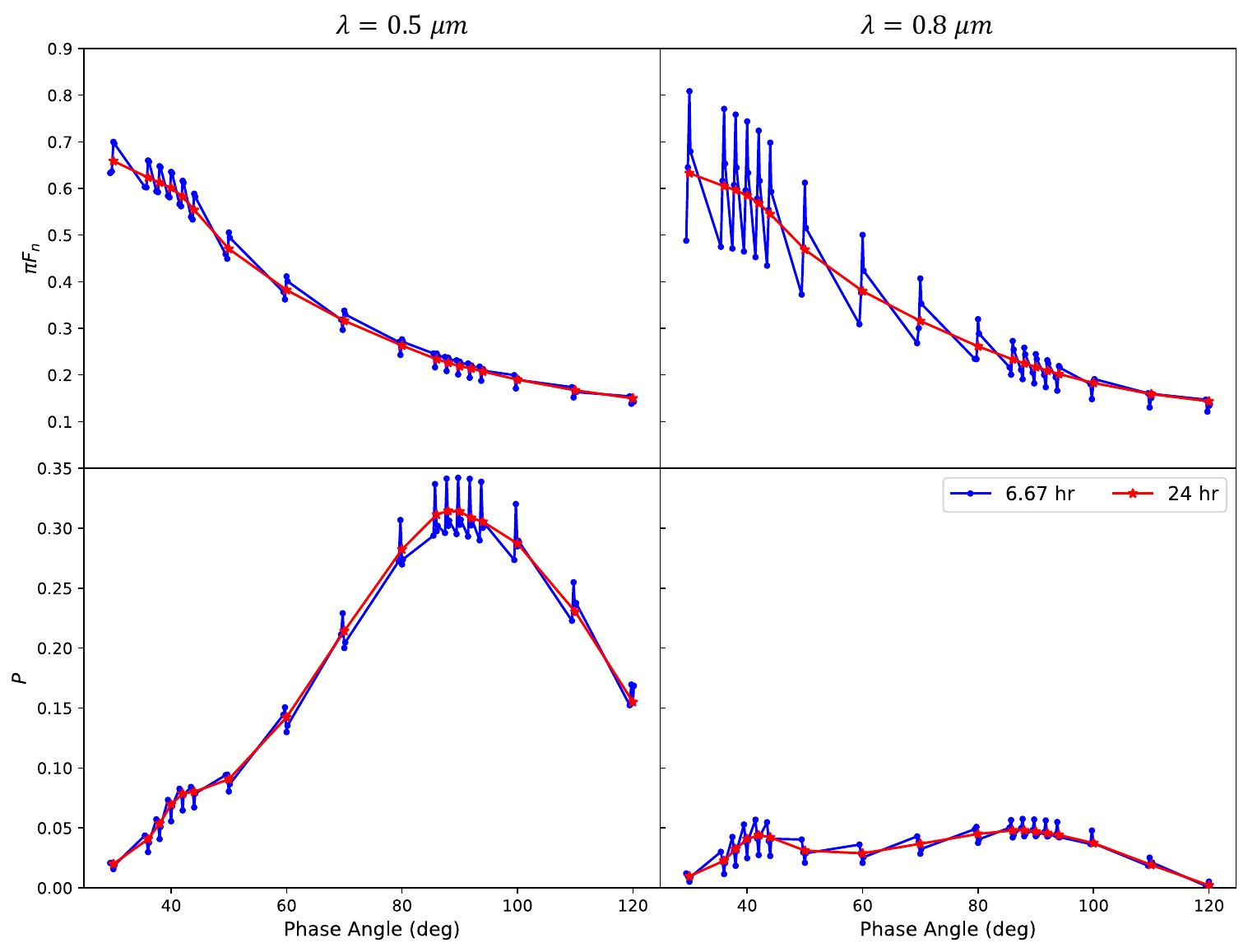}
    \caption{$\pi{F_{n}}(\alpha)$ (top) and $P(\alpha)$ (bottom) for $\lambda = 0.5$ $\mu$m (left) or $\lambda = 0.8$ $\mu$m (right) for our modern Earth of Fig.~\ref{fig: ModEarthviews}. The planet rotates on its axis once every 24 hr. Variability due to the different surfaces and clouds moving into and out of view lead to the ``noise" observed in the 6.67 hr observations (blue circles), while these variable features get smoothed out in the 24 hr observations (red stars).}
    \label{fig: ModEarthInt}
\end{figure*}

Fig.~\ref{fig: ModEarthInt} shows the $\pi{F_{n}}$ (top) and $P$ (bottom) phase curves for our modern Earth model, including the effect of diurnal rotation (blue circles). Our model planet was observed at either $\lambda = 0.5$ $\mu$m (left) or $\lambda = 0.8$ $\mu$m (right). When observed at time steps of $\sim$6.67 hr (blue circles), the different clouds and surfaces rotating into and out of view due to the diurnal rotation result in a ``noisy'' signal. Once integrated over 24 hr time steps (red stars), the features of the diurnal variation are smoothed out. In  $\pi{F_{n}}$, the brightness of the planet shows a clear dimming from short to long phases, and the $\pi{F_{n}}$ curves are almost identical between the two wavelengths. Meanwhile, the overall $P$ signals retain the key markings of the modern Earth atmosphere; particularly, the Rayleigh scattering peak expected at $\sim$90$\degree$ and the rainbow peak of the liquid water clouds at $\sim$40$\degree$. The $P$ curves also display a clear wavelength dependence, making them valuable tools for characterizing our modern Earth scenario.

\begin{figure*}[ht!]
    \centering
    \includegraphics[width=11.5cm]{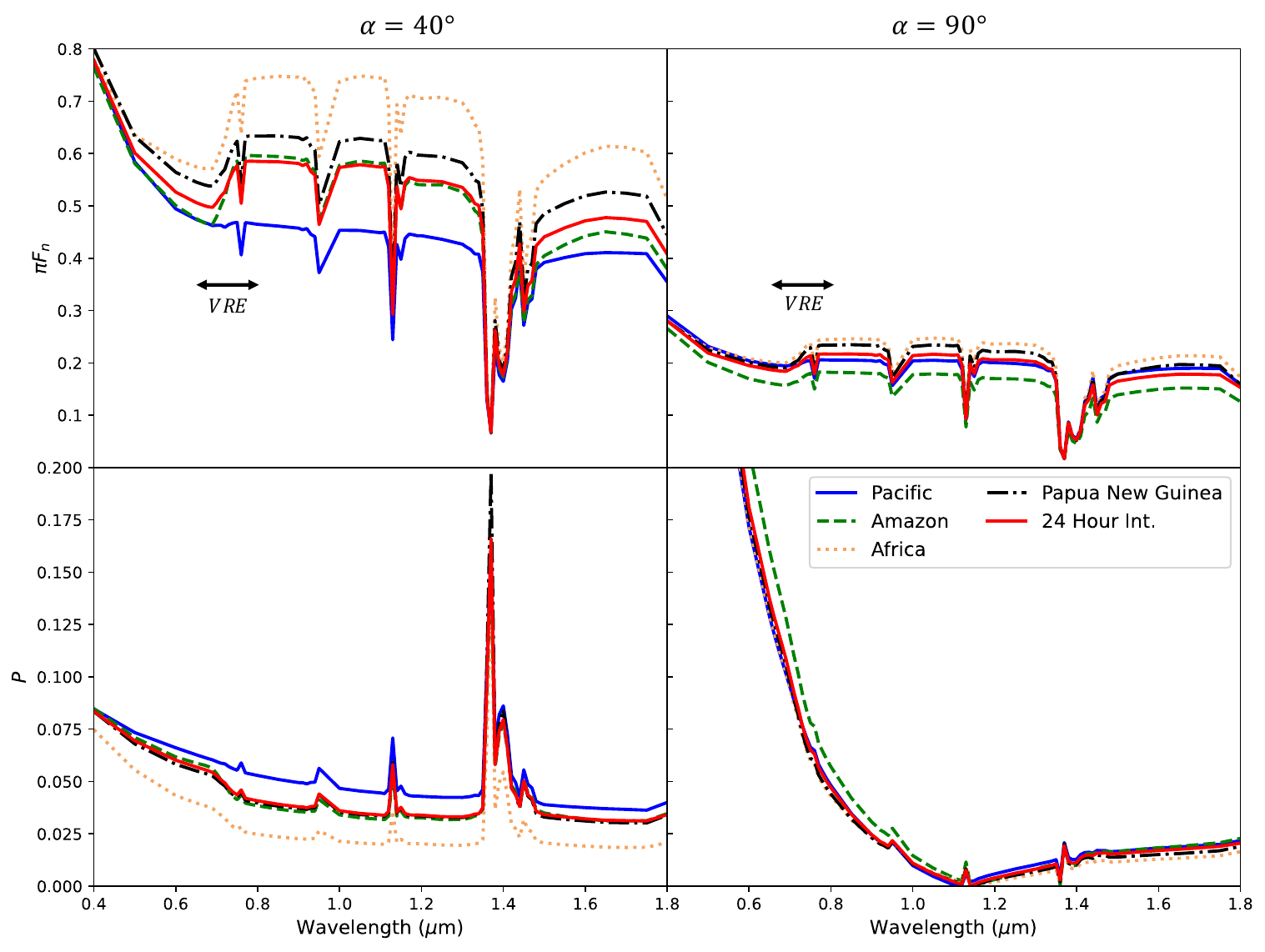}
    \caption{$\pi{F_{n}}(\lambda)$ (top) and $P(\lambda)$ (bottom) for $\alpha = 40\degree$ (left) or $\alpha = 90\degree$ (right) for the four separate views of modern Earth displayed in Fig.~\ref{fig: ModEarthviews}. At $\alpha = 90\degree$, the polarization of all 4 models and the 24-hour integration reach a maximum value of $\sim$0.5 at $\lambda = 0.4$ $\mu$m. While the $\pi{F_{n}}(\lambda)$ curves show larger separations in the signals between the different views, the $P(\lambda)$ curves are affected more by the heterogeneities.}
    \label{fig: ModEarthspecs}
\end{figure*}

Fig.~\ref{fig: ModEarthspecs} shows the $\pi{F_{n}}$ (top) and $P$ 
spectra for the modern Earth as seen at each of the four views of Fig.~\ref{fig: ModEarthviews}, observed at either $\alpha = 40\degree$ (left) or $\alpha = 90\degree$ (right). Also included are spectra of the modern Earth integrated over a full 24 hour diurnal rotation (red lines). At $\alpha = 90\degree$, the presence of water clouds results in a reduction of the depths of the gaseous absorption bands, especially in $P$ (bottom right), and the models become harder to separate compared to clear atmosphere models \citep[e.g., ][their Figs.~13 and 15]{goodisgordon2025}. The Rayleigh scattering results in high levels of $P$ at shorter $\lambda$, up to values of $\sim$0.5 at $\lambda = 0.4$ $\mu$m. At $\alpha = 40\degree$, near the peak of the rainbow feature, there is better separation between the four views in both $\pi{F_{n}}$ and $P$. The highly reflective sand dominating the Africa view (gold dotted line) produces larger $\pi{F_{n}}$ than the other models where forest or ocean dominate the surface. More interestingly, however, we see that the distributions of the vegetated surfaces on the planetary disk have larger impacts on the resulting $P$ than they do on $\pi{F_{n}}$. Whereas the $\pi{F_{n}}$ curves have very similar vegetation red edge (VRE; i.e., sharp change in reflectivity around 0.7 $\mu$m due to photosynthetic organisms) signals for both the Africa (dotted gold lines) and Amazon (dashed green lines) views, with both increasing by $\sim$13\%, the $P$ curves show a sharper VRE for the Amazon view than the Africa view, with $P$ decreasing by $\sim$1.5\% for the Amazon view but $<$1\% for the Africa view. This is because the vegetation for the Amazon view is centered on the observable disk, thus playing a larger role in the resulting disk-integrated polarization of the planet. Observing the planet over a full 24-hour integration at either phase angle results in smaller but still prominent VRE signals in both $\pi{F_{n}}$ and $P$, indicating that the habitability signs of the vegetation are not lost due to diurnal rotation.

\subsection{Surface Distributions} \label{sec: SurfDist}

\begin{figure}[ht!]
    \centering
    \includegraphics[width=\linewidth]{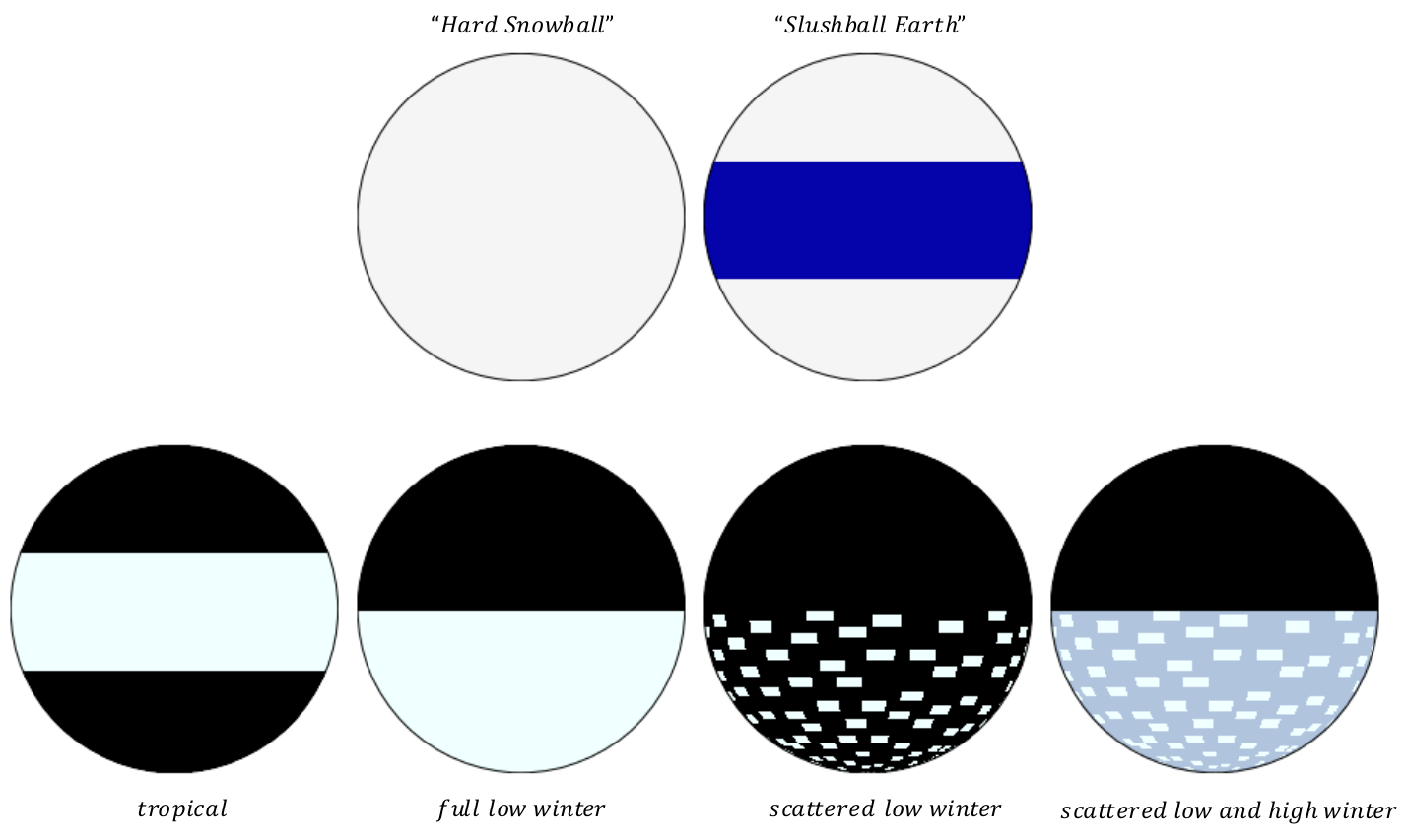}
    \caption{Surface (top row) and cloud (bottom row) distributions for our various Snowball Earth models. The surface was either modeled as a fully ice-covered ``Hard Snowball" (top left) or contained a tropical ocean belt to simulate a ``Slushball Earth" scenario (top right). Low-altitude water ice clouds (light blue) could have existed in an equatorial belt during summer months (bottom left) or only in the southern hemisphere during winter months. In winter these low-level clouds could have been fully covering the hemisphere (bottom middle left), scattered throughout the hemisphere (bottom middle right), or scattered throughout with additional high-altitude (grayish-blue) water ice clouds (bottom right).}
    \label{fig: SnowEarthviews}
\end{figure}

The early Earth is known to have gone through a few different periods of global glaciation, resulting in multiple Snowball Earth events. However, a debate still exists on whether the planet was entirely covered by thick ice sheets, termed a ``Hard Snowball'' \citep[hereafter HS; e.g.,][]{kirschvink1992, hoffman1998}, or if there existed a tropical ocean belt around the equator to aid with deglaciation, termed a ``Slushball Earth'' \citep[hereafter SE; e.g.,][]{hyde2000, abbot2011}. Here we show the effects of including a band of open ocean on a Snowball Earth on the resulting $\pi{F_{n}}$ and $P$ signals of the planet. The top row of Fig.~\ref{fig: SnowEarthviews} shows the two different surface distributions for our HS and SE scenarios. Each of these scenarios could have possessed various cloud distributions across their surfaces, which are discussed in Sect.~\ref{sec: Weather}. In this section, we stick to showing the differences between the two scenarios with either clear atmospheres or atmospheres containing an equatorial belt of optically thick ($\tau$ = 10) water ice clouds (bottom left of Fig.~\ref{fig: SnowEarthviews}). These ice clouds are expected to have resembled low-level stratocumulus clouds on the modern Earth \citep[e.g.,][]{abbot2014} and are therefore placed at an altitude of $\sim$1 km.

\begin{figure*}[ht!]
    \centering
    \includegraphics[width=11.5cm]{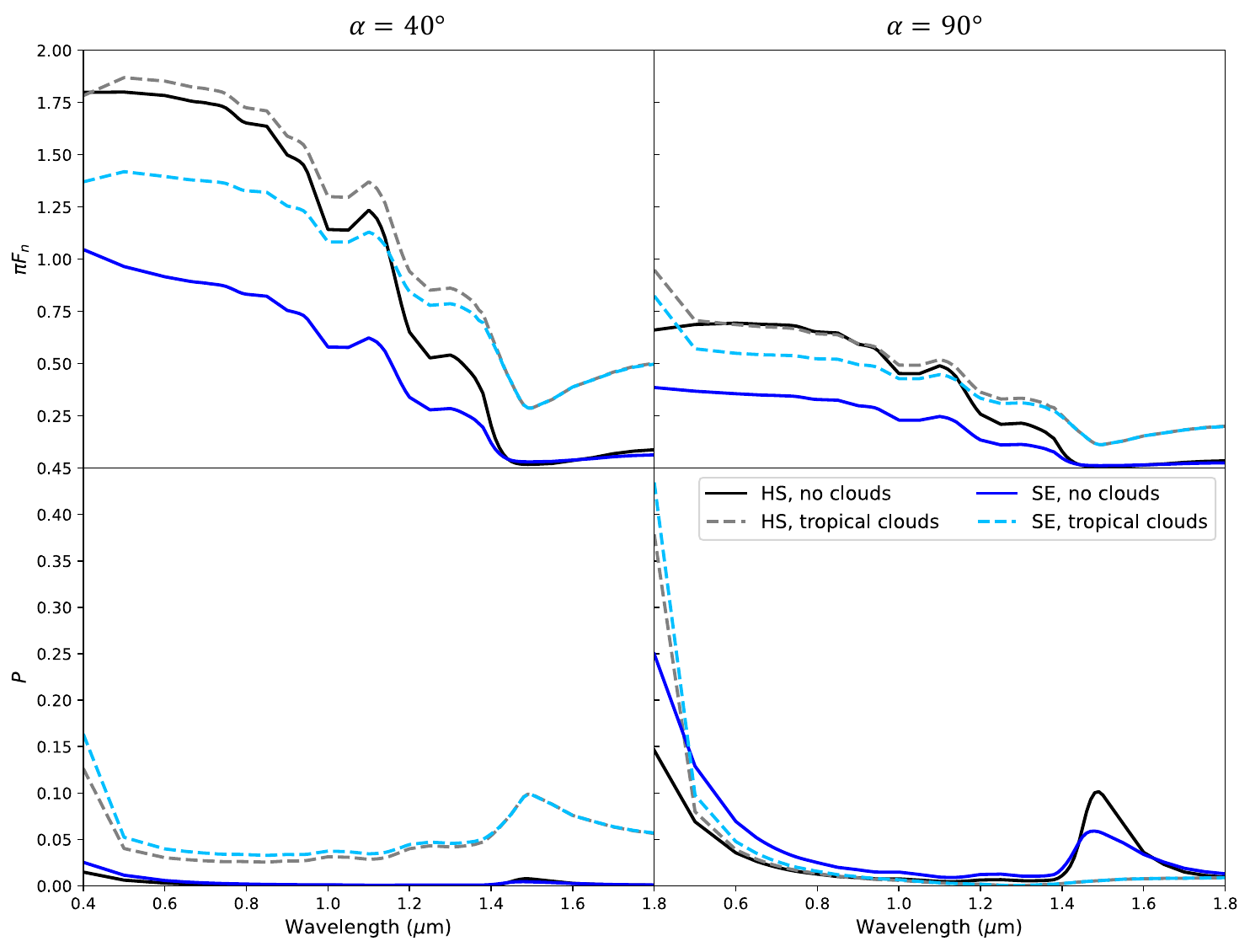}
    \caption{$\pi{F_{n}}(\lambda)$ (top) and $P(\lambda)$ (bottom) for $\alpha = 40\degree$ (left) or $\alpha = 90\degree$ (right) for our ``Hard Snowball" (HS) and ``Slushball Earth" (SE) scenarios with either clear atmospheres (solid lines) or atmospheres containing a band of optically thick water ice clouds around the equator (dashed lines). Reflection and scattering by the clouds increase both the reflectivity and polarization of the planets at $\alpha = 40\degree$, while at $\alpha = 90\degree$ the clouds increase $\pi{F_{n}}$ but decrease $P$.}
    \label{fig: SnowSlushwaves}
\end{figure*}

Fig.~\ref{fig: SnowSlushwaves} displays the resulting unpolarized (top) and polarized (bottom) spectra of our HS (black) and SE (blue) scenarios with either clear (solid lines) or cloudy (dashed lines) atmospheres. The planets are observed at either $\alpha = 40\degree$ (left) or $\alpha = 90\degree$ (right). As expected, the introduction of a dark ocean belt on the center of the observable disk decreases the reflectivity of the planet, resulting in lower $\pi{F_{n}}$ for the SE scenario in both the clear and cloudy cases. On the other hand, the reflectivity of the icy surface coupled with the homogeneity of the HS scenario results in nearly zero $P$, especially at $\alpha = 40\degree$. The addition of the tropical ocean breaks this homogeneity for the SE scenario, but while this causes separation of the signals in $P$ at $\alpha = 90\degree$, the two scenarios remain indistinguishable at $\alpha = 40\degree$ due to the large ice surface coverage at this phase. The increased reflection from the addition of the ice particles in the water clouds causes a higher $\pi{F_{n}}$ for the cloudy cases in both scenarios. Similarly, scattering of the light from the water ice clouds introduces additional heterogeneities into the resulting planetary signals and increases the resulting $P$ for the planet at $\alpha = 40\degree$, mainly due to the presence of the water cloud rainbow feature around this phase angle. However, increased multiple scattering within the cloud particles at $\alpha = 90\degree$ actually results in a net decrease in the polarization of the planet at this phase angle.

\begin{figure}[ht!]
    \centering
    \includegraphics[width=\linewidth]{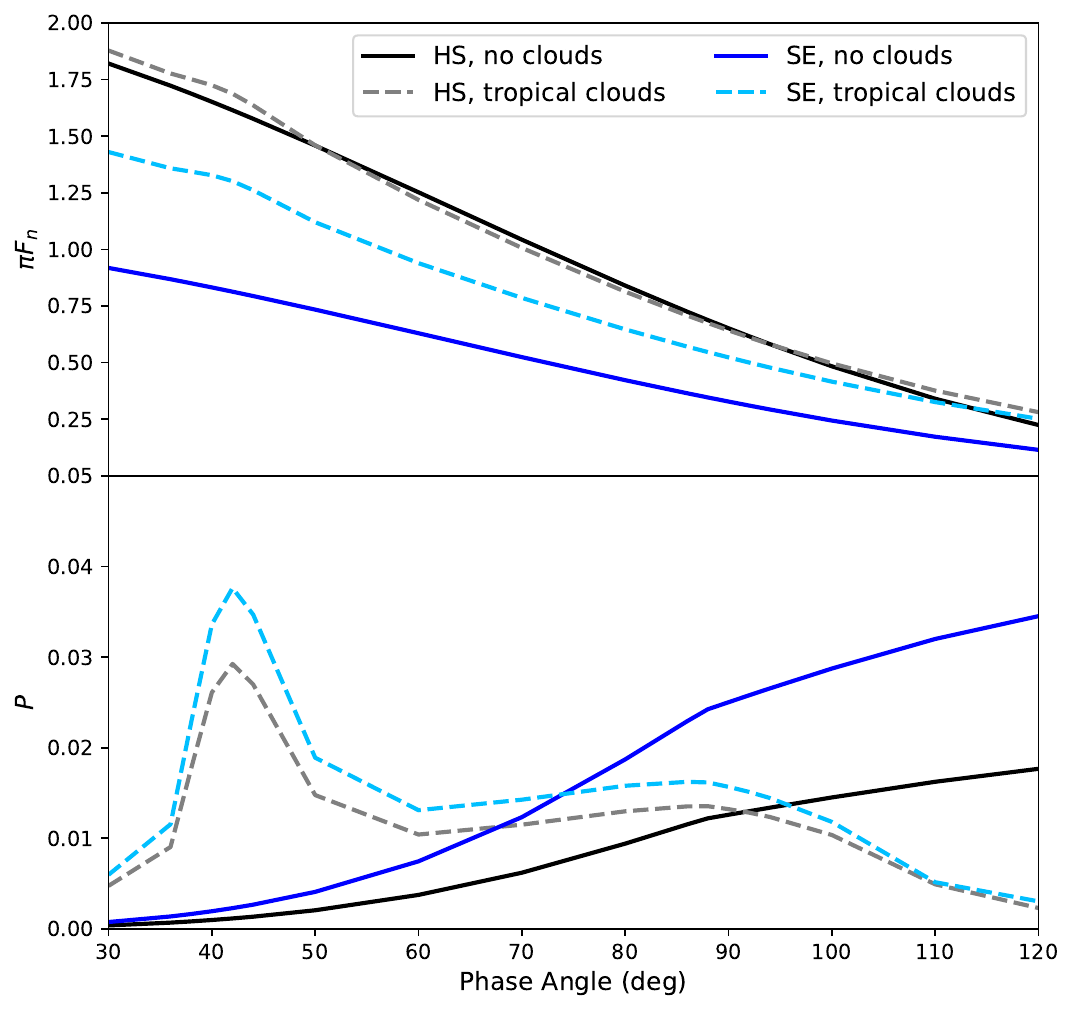}
    \caption{$\pi{F_{n}}(\alpha)$ (top) and $P(\alpha)$ (bottom) for $\lambda = 0.8$ $\mu$m for our ``Hard Snowball" (HS) and ``Slushball Earth" (SE) scenarios with either clear atmospheres (solid lines) or atmospheres containing a band of optically thick water ice clouds around the equator (dashed lines). The peaks in $P$ around $\alpha=40\degree$ are due to modeling our water ice cloud particles using Mie theory.}
    \label{fig: SnowSlushphases}
\end{figure}

Fig.~\ref{fig: SnowSlushphases} shows $\pi{F_{n}}(\alpha)$ (top) and $P(\alpha)$ (bottom) for the HS (black) and SE (blue) scenarios with either clear (solid lines) or cloudy (dashed lines) atmospheres, observed at $\lambda = 0.8$ $\mu$m. As expected, the reflectivity of the planet decreases with increasing $\alpha$ for all cases as more of the observable disk is in the nightside. The addition of the water ice clouds increases $\pi{F_{n}}$ for the SE scenario but has only a very small effect on the signals for the HS scenario due to the symmetry of the underlying homogeneous ice surface. However, in $P$ clear differences can be detected between all four cases. For the clear atmosphere cases, the symmetry of the planetary surface combined with the Rayleigh scattering by the atmosphere causes negligible $P$ at the shortest $\alpha$ but larger $P$ at the longer $\alpha$, as the Rayleigh peaks for these cases are pushed off toward $\alpha = 135\degree$. The water ice clouds introduce an additional rainbow peak in the $P$ curves around $\alpha = 40\degree$ for both scenarios, which is due to our assumption that the water ice cloud particles are spherical. We acknowledge that natural water ice cloud particles are nonspherical and so we would not expect to see these peaks in the phase curves of planets with more realistic water ice clouds \citep[e.g.,][and references therein]{goodisgordon2025}. However, modeling water ice clouds with nonspherical particles is outside the scope of this study.

\subsection{Seasonal Weather Patterns} \label{sec: Weather}

While most studies agree that clouds were necessary in Snowball Earth events to aid in deglaciation \citep[e.g.,][]{abbot2012, abbot2014, braun2022}, little is known about the true properties or distributions of water clouds during these early Earth events. \citet[][]{abbot2011} presented evidence that a specific type of SE event, termed the ``Jormungand” state, could have allowed clouds to form over a thin strip of open ocean near the equator and then dissipate starting in the subtropics. \citet[][]{braun2022} argued that the uncertainties inherent in mixed-phase (i.e., liquid water and water ice) clouds used in their SE models made it difficult to determine the true properties of the climate and clouds during these events. \citet[][]{abbot2014} used GCM and cloud-resolving models to simulate a HS Earth and found that the models produced low-level, optically thick water ice clouds, but these simulations did not provide any indication of the global distribution of these clouds. Recently, \citet[][]{yan2024} used high-resolution GCM models to simulate clouds on a HS Earth and found that the clouds showed clear variability between different parts of the orbit, with a band of clouds over the equator during Northern Hemisphere (NH) summer months and a hemispherical distribution of clouds throughout the southern hemisphere (SH) during NH winter months.

Here we investigated the effects that different distributions of water ice clouds, representing different seasons, have on the resulting $\pi{F_n}$ and $P$ signals of a Snowball Earth. The bottom row of Fig.~\ref{fig: SnowEarthviews} displays four different distributions of water ice clouds we considered here: a tropical belt of low-altitude clouds for the summer months (bottom left), and three different cases for the winter months: a blanket of low-altitude clouds covering the entirety of the SH (bottom middle left); randomly scattered low-altitude clouds throughout the SH (bottom middle right); or randomly scattered low-altitude clouds with the remaining pixels filled by higher-altitude ice clouds (bottom right). The low-altitude clouds all had $\tau$ = 10 and were placed at $\sim$1 km in the atmosphere, while the higher-altitude clouds all had $\tau$ = 5 and were placed at $\sim$4.5 km in the atmosphere.

\begin{figure*}[ht!]
    \centering
    \includegraphics[width=11.5cm]{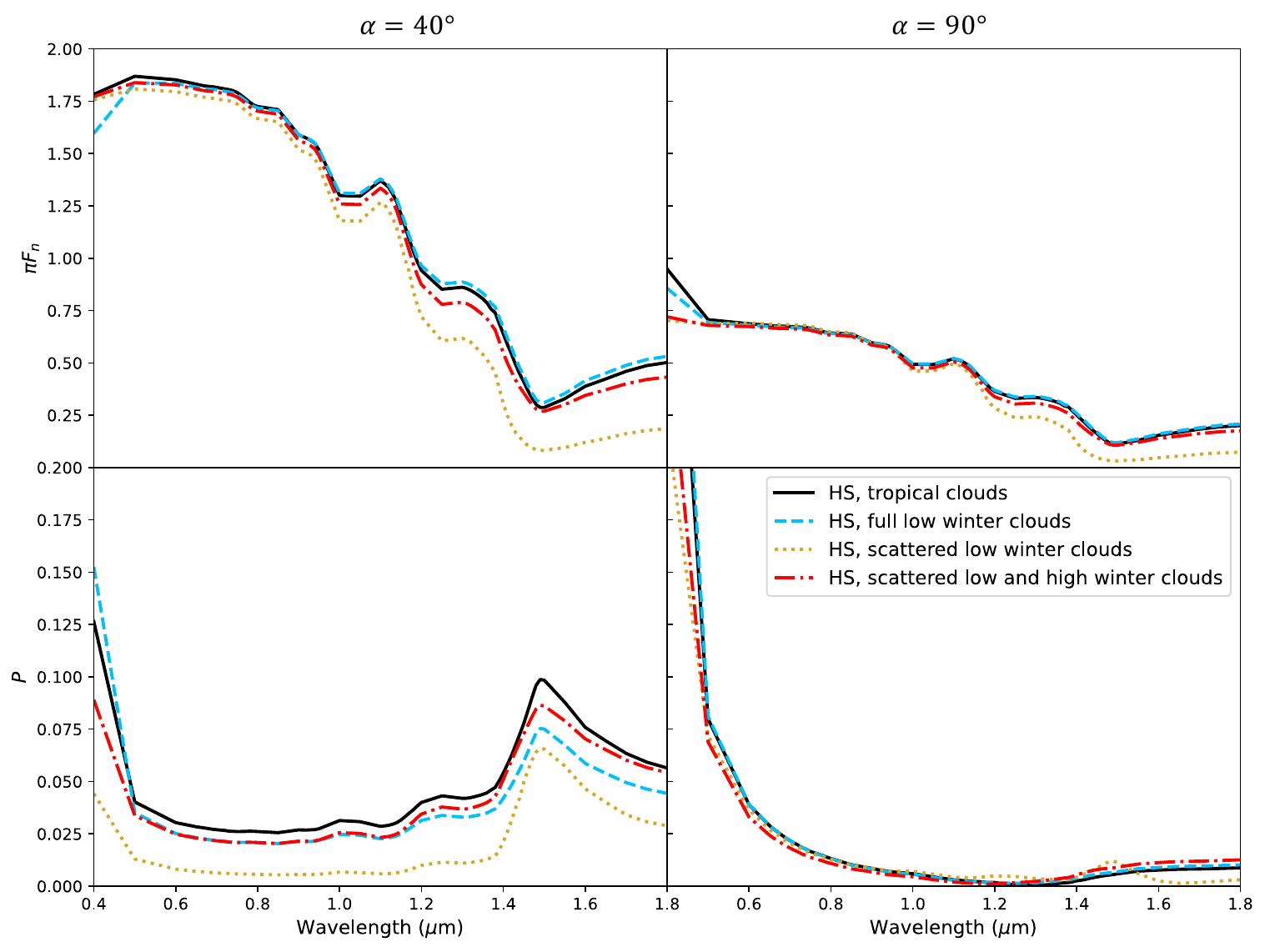}
    \caption{$\pi{F_{n}}(\lambda)$ (top) and $P(\lambda)$ (bottom) for $\alpha = 40\degree$ (left) or $\alpha = 90\degree$ (right) for our ``Hard Snowball" (HS) scenario with various distributions of water ice clouds. The $P$ curves show clearer separation than the $\pi{F_{n}}$ curves, especially at $\alpha = 40\degree$.}
    \label{fig: SnowWinSumSpec}
\end{figure*}

Fig.~\ref{fig: SnowWinSumSpec} shows $\pi{F_n}(\lambda)$ and $P(\lambda)$ of the four HS cases observed at either $\alpha = 40\degree$ (left) or $\alpha = 90\degree$ (right). Clouds condense most of the spectra together, causing overlap between the spectra that makes it difficult to differentiate the various cloud distributions, especially when the planet is observed at $\alpha = 90\degree$. When observed at $\alpha = 40\degree$, however, the $P(\lambda)$ spectra show larger separations than the $\pi{F_{n}}(\lambda)$ spectra, especially for the case of scattered low-altitude clouds (dotted gold lines). The combination of cloudy and clear pixels from this case lowers the continuum by up to $\sim$0.03 compared to the other cases in which the entire SH is fully cloud covered.

\begin{figure}[ht]
    \centering
    \includegraphics[width=\linewidth]{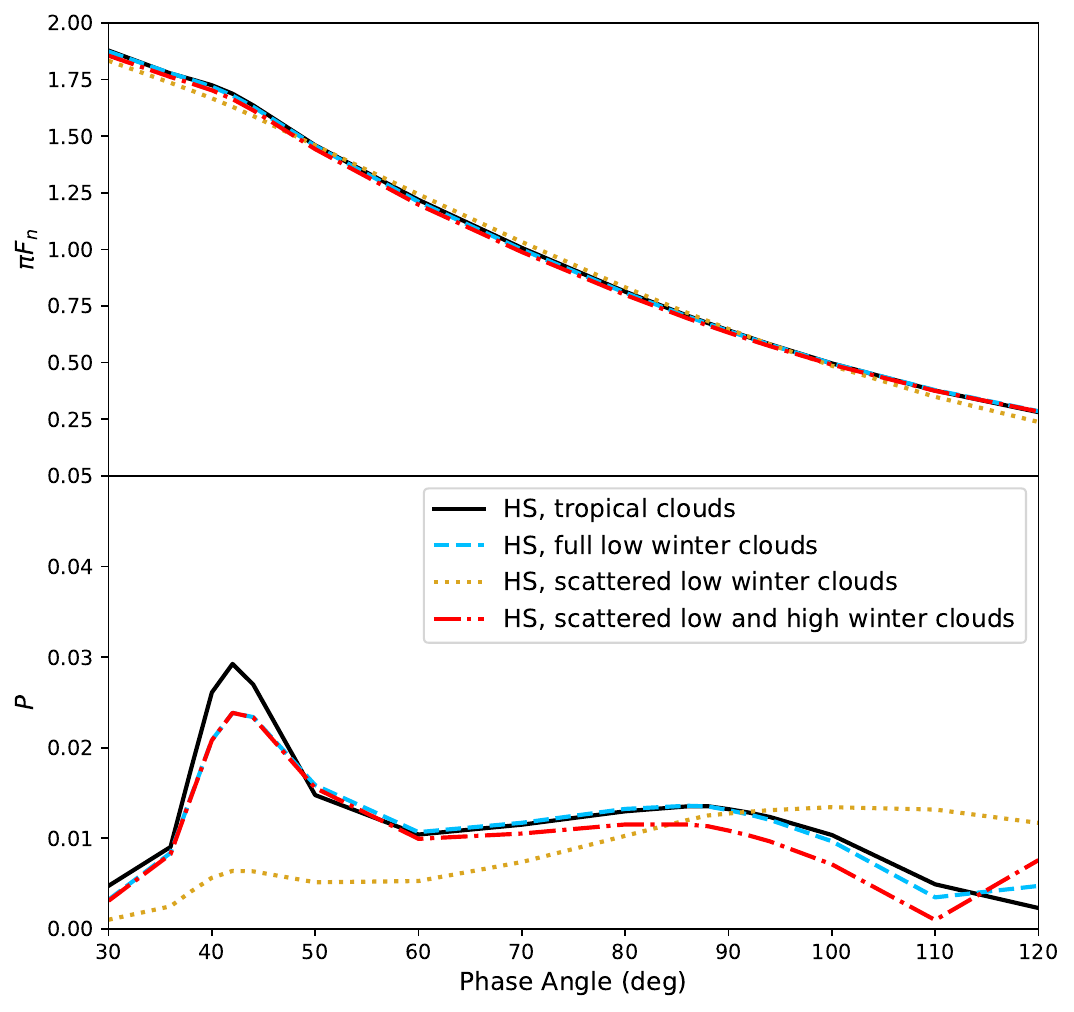}
    \caption{$\pi{F_{n}}(\alpha)$ (top) and $P(\alpha)$ (bottom) for $\lambda = 0.8$ $\mu$m for our ``Hard Snowball" (HS) scenario with various distributions of water ice clouds. While the flux curves overlap in all cases, the polarized curves show clear differences between the varying distributions.}
    \label{fig: SnowWinSumPhase}
\end{figure}

In Fig.~\ref{fig: SnowWinSumPhase} we show the resulting $\pi{F_{n}}(\alpha)$ (top) and $P(\alpha)$ (bottom) for the four HS cases observed at $\lambda = 0.8$ $\mu$m. Similar to Fig.~\ref{fig: SnowWinSumSpec}, the flux phase curves show a significant overlap among the four cases, with negligible differences between the phase curves. However, we can now see some variability between the signals of the four cases in the polarized phase curves. The placement of the tropical clouds (solid black lines) at the center of the observable disk has a stronger influence on the overall disk-integrated $P$, and therefore results in a larger rainbow peak at $\alpha \sim 40\degree$ than the three models with the clouds all in the SH. Although the full lower clouds model (dashed blue lines) and the mixed low and high clouds model (dashed dotted red lines) show more overlap, the fact that some of the clouds are higher in the atmosphere in the mixed case leads to a smaller Rayleigh peak near $\alpha = 90\degree$ than the full lower clouds case, as the light now interacts less with the atmospheric gases and more with the higher--altitude clouds. Finally, the mix of clear with cloudy atmosphere pixels in the scattered clouds case (gold dotted lines) results in less light interacting with cloud particles, thus lowering the rainbow peak and raising the Rayleigh peak. These models thus highlight the crucial, and complimentary to flux, role that polarization plays in characterizing the distributions of features on the observed planet.

Up until now, our models used static clouds. In reality, atmospheric circulation causes clouds to move across the planet throughout the course of one day, as well as grow, dissipate, and evolve over the course of different seasons throughout Earth's orbit. We now show the effect that atmospheric circulation and seasonal variations have on the resulting signals of our modern Earth scenario.

\begin{figure*}[ht]
    \centering
    \includegraphics[width=11.5cm]{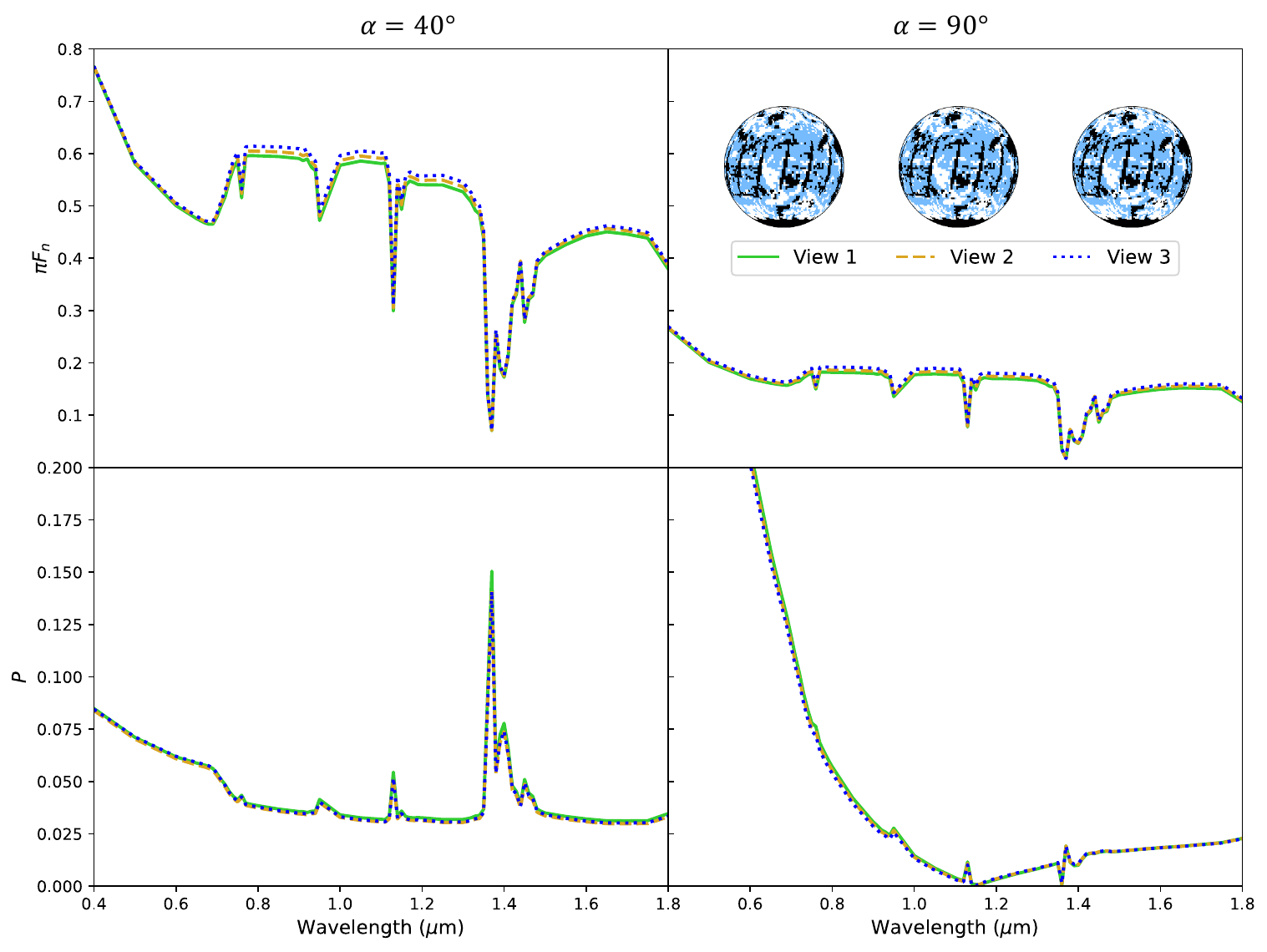}
    \caption{$\pi{F_{n}}(\lambda)$ (top) and $P(\lambda)$ (bottom) for $\alpha = 40\degree$ (left) or $\alpha = 90\degree$ (right) for our modern Earth scenario, centered on the Amazon, with three different views of the clouds rotating throughout one day. The distributions of the clouds for each view can be seen in the inlaid globe plots, where View 1 corresponds to the original Amazon cloud distribution of Fig.~\ref{fig: ModEarthviews}.}
    \label{fig: ModAmazAtmCircSpec}
\end{figure*}

\begin{figure}[ht]
    \centering
    \includegraphics[width=\linewidth]{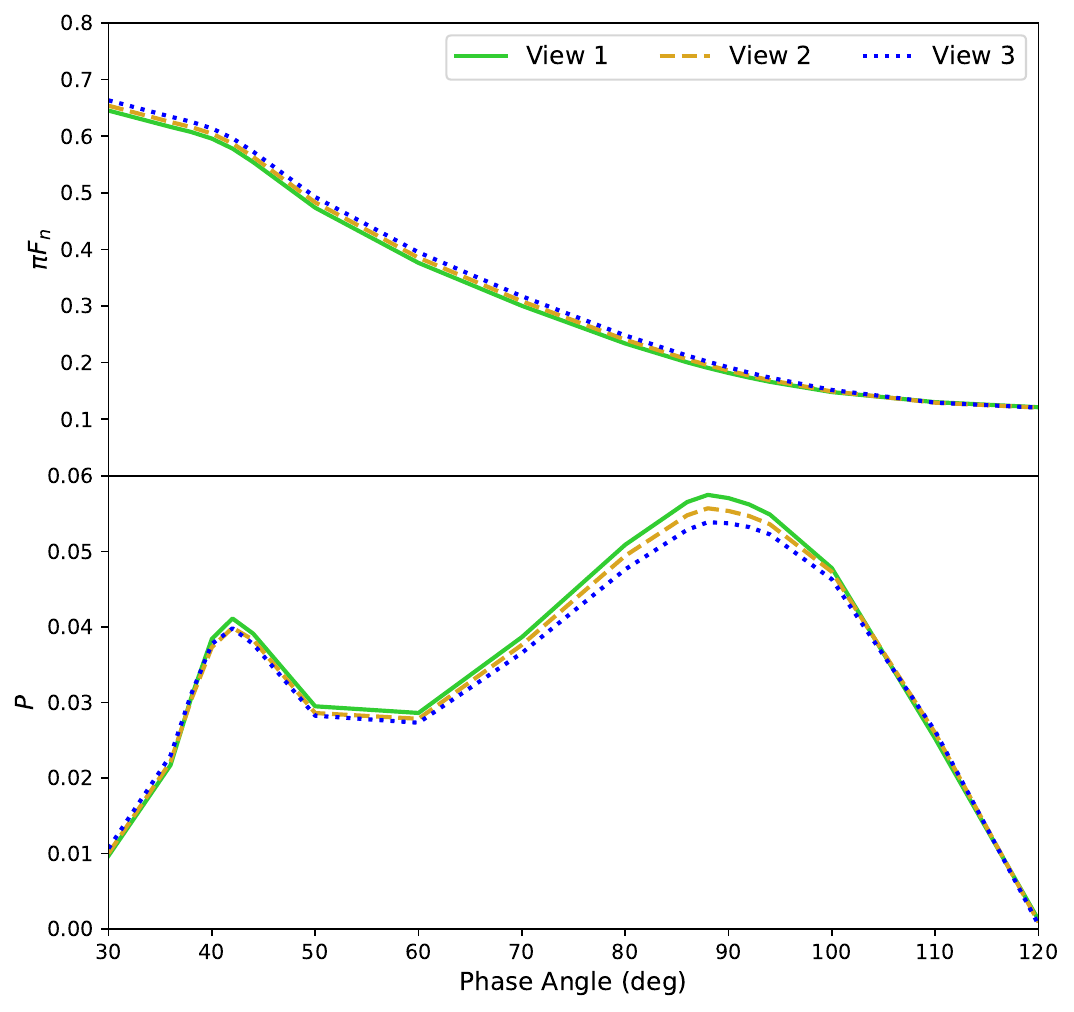}
    \caption{$\pi{F_{n}}(\alpha)$ (top) and $P(\alpha)$ (bottom) for $\lambda = 0.8$ $\mu$m for our modern Earth scenario, centered on the Amazon, with three different views of the clouds rotating throughout one day. Small variations between the different views can be seen in both the flux and polarized phase curves.}
    \label{fig: ModAmazAtmCircPhase}
\end{figure}

We simulated modern Earth with the Amazon centered on the observable disk (see middle left of Fig.~\ref{fig: ModEarthviews}) and allowed the clouds to rotate randomly over the course of one day. The clouds moved at a rate of $\sim$50 km/hr, which corresponds to the average speed for low-level clouds in the Earth's atmosphere \citep[e.g.,][]{hasler1976}.
Fig.~\ref{fig: ModAmazAtmCircSpec} shows the resulting $\pi{F_n}(\lambda)$ and $P(\lambda)$ for three different views of these rotating clouds taken $\sim$8.5 hours apart (see inlet maps). The planet is shown again observed at either $\alpha = 40\degree$ (left) or $\alpha = 90\degree$ (right). While there are some small yet detectable variations in $\pi{F_{n}}$ between the different views at both phases, all of the signals in $P$ overlap and show negligible differences. Fig.~\ref{fig: ModAmazAtmCircPhase} shows $\pi{F_{n}}(\alpha)$ (top) and $P(\alpha)$ (bottom) at $\lambda = 0.8$ $\mu$m for the three different cloud views. While small differences can be detected between the signals in $P$, such as a slightly lower Rayleigh peak ($\Delta{P} \approx 0.0035$) in View 3 due to more clear atmosphere pixels rotating into view compared to View 1, the overall trends and shapes of the phase curves remain the same and we can still see more defining features than those in the $\pi{F_{n}}$ curves. These results suggest that the polarization is less sensitive to the ``noise" introduced by these small-scale weather changes, thus making it a more robust tool than flux in characterizing our modern Earth.

\begin{figure*}[ht]
    \centering
    \includegraphics[width=11.5cm]{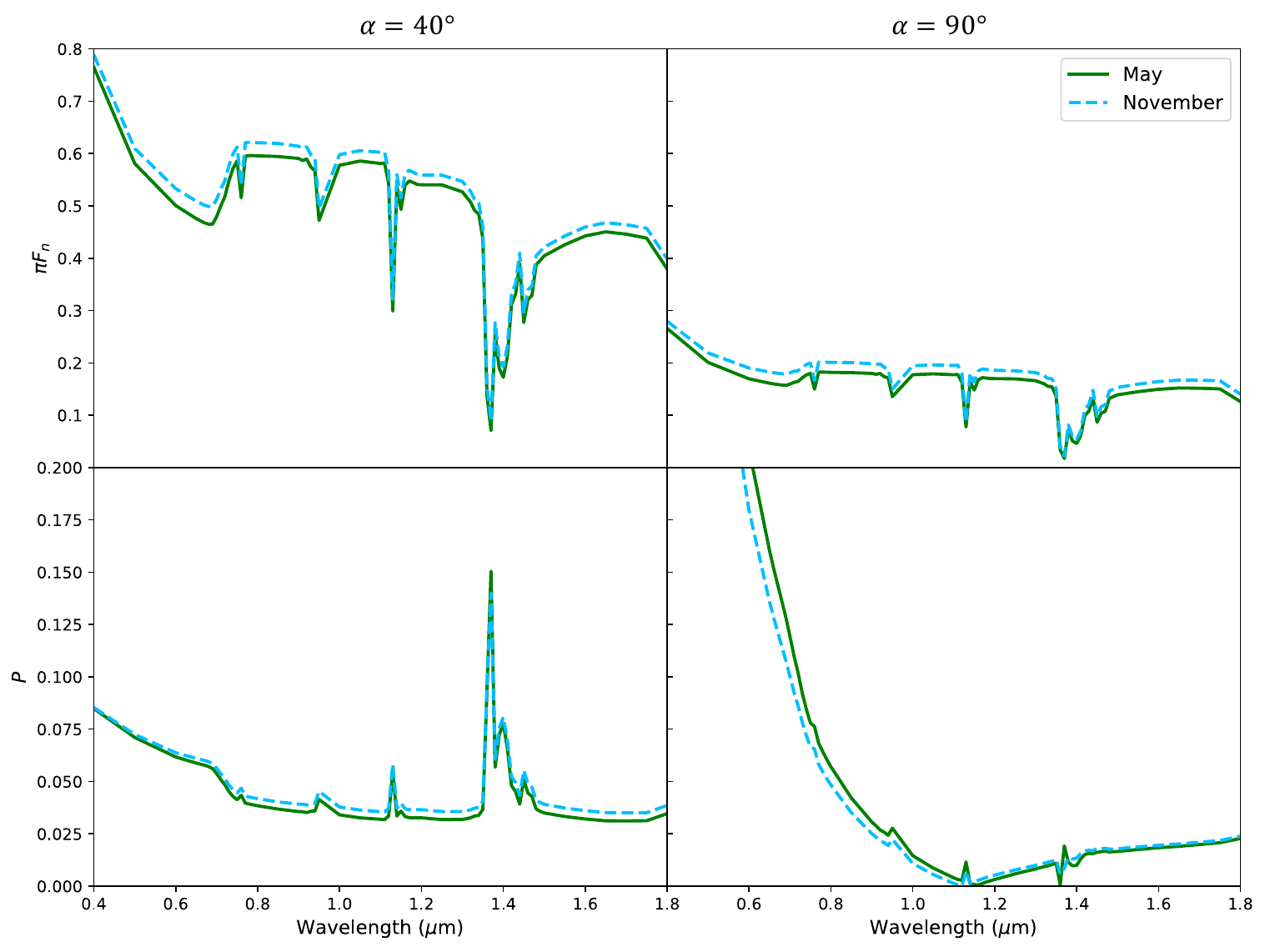}
    \caption{$\pi{F_{n}}(\lambda)$ (top) and $P(\lambda)$ (bottom) for $\alpha = 40\degree$ (left) or $\alpha = 90\degree$ (right) for our modern Earth scenario, centered on the Amazon, with surface and cloud distributions taken from MODIS data observed in May (solid green lines) or November (dashed blue lines).}
    \label{fig: ModWinSumSpec}
\end{figure*}

Lastly, we simulated the modern Earth with the Amazon still dominating the view but now for two different points in the planetary orbit, either late NH spring or late NH autumn. Specifically, we used MODIS land and cloud data obtained for 18 May 2013 (matching the Amazon view of our previous models) and 18 November 2013 (6 months difference). In the latter case, the snow and ice surface coverage in the NH increased significantly, reaching latitudes as low as 45$\degree$ N, while the Amazon rainforest showed a darker average albedo but still a decent amount of cloud coverage as it was near the transition from the tropical dry season to the tropical wet season. The resulting $\pi{F_n}$ (top row) and $P$ (bottom row) spectra for these two seasonal views are shown in Fig.~\ref{fig: ModWinSumSpec} for $\alpha = 40\degree$ (left column) or $\alpha = 90\degree$ (right column). Similarly to the different modern Earth views of Fig.~\ref{fig: ModEarthspecs}, the differences in the surface distributions between the two seasonal views lead to slight changes in the VRE signal, especially at $\alpha = 40\degree$. In particular, while the $\pi{F_n}$ of both views increase by $\sim$12\% at this phase, the $P$ of the May case (solid green lines) decreases by $\sim$2\% while the $P$ of the November case (dashed blue lines) decreases by $\sim$1.8\%. These small differences are due to the snow coverage being located near the limbs of the observable disk rather than the center, therefore contributing less to the total disk-integrated $P$. On the other hand, the increased snow coverage clearly increases the reflectivity of the planet in the November case, leading to a higher continuum in $\pi{F_n}$, especially at shorter $\lambda$ due to the spectral shape of the ice albedo. While both $\alpha$'s show some separations at various $\lambda$ between the spectra for both cases, the presence of the water and ice clouds across the globe cause some overlap of the signals, especially at $\alpha = 90\degree$, even though the distributions of the clouds themselves are changing. Our results are in line with previous analyses of seasonal variations in modern Earth signatures from \citet[][]{roccetti2025}.

\subsection{Early versus Modern Mars} \label{sec: EarlyModMars}

Mars has received significant attention over the last century and has thus been observed with polarimetry by numerous studies \citep[see, e.g., overviews by][]{ebisawa1993, shkuratov2005, bagnulo2024}. Due to the geometry of the solar system, Mars can only be observed by ground-based observatories to a maximum phase angle of $\alpha \approx 45\degree$. Intermediate phase angle polarimetric observations were only achieved with the VPM photopolarimeters on the USSR's MARS-5 spacecraft in orbit around the planet \citep[see, e.g.,][]{santer1985}.

\begin{figure}[ht]
    \centering
    \includegraphics[width=\linewidth]{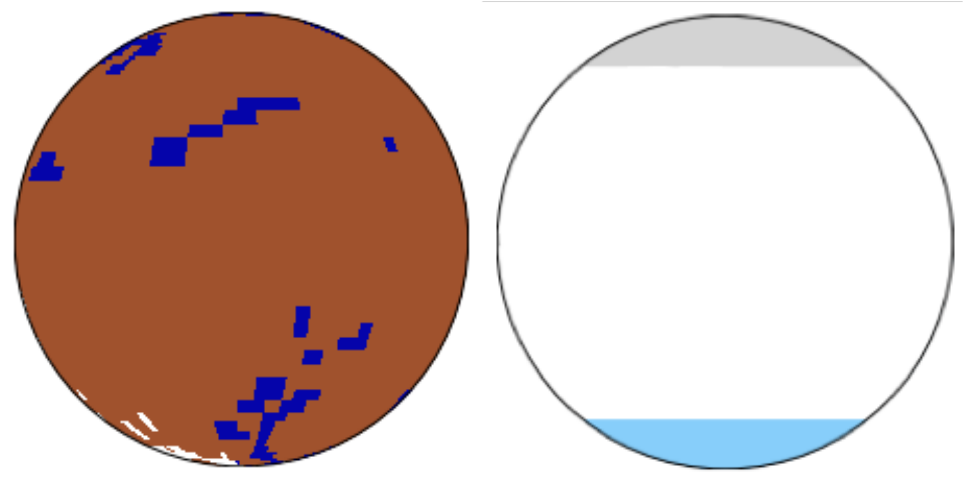}
    \caption{Surface (left) and aerosol (right) distributions for our early Mars simulation. The surface was comprised of bare soil (brown), open water (blue), and sea/lake ice and snow (white) based on ROCKE-3D calculations. The properties of the water ice clouds on early Mars were computed for three distinct latitudinal regions (see Sect.~\ref{sec: EarlyMars}): a North Polar (NP) region (gray), an Equatorial (EQ) region (white), and a South Polar (SP) region (blue).}
    \label{fig: Marsviews}
\end{figure}

\begin{figure}[ht]
    \centering
    \includegraphics[width=\linewidth]{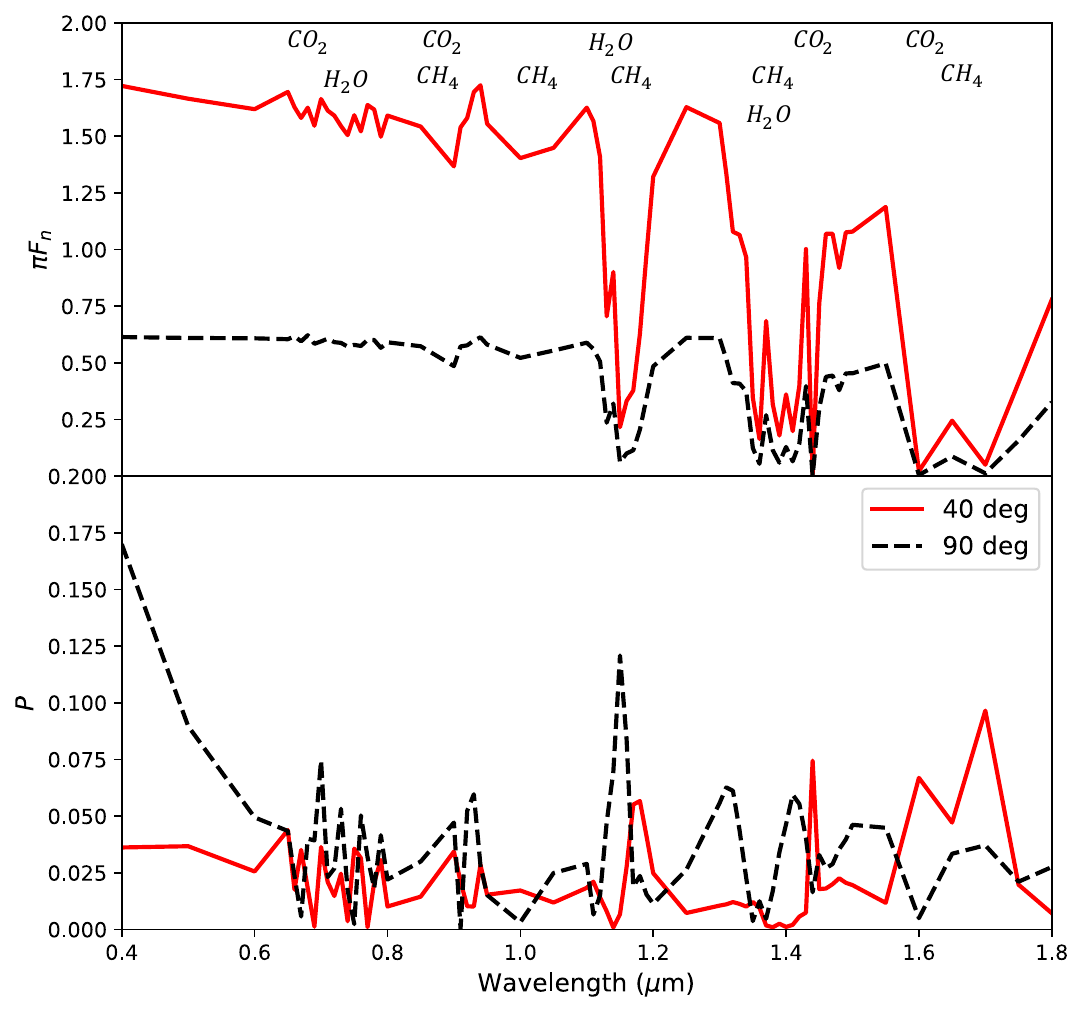}
    \caption{$\pi{F_{n}}(\lambda)$ (top) and $P(\lambda)$ (bottom) for $\alpha = 40\degree$ (solid red lines) or $90\degree$ (dashed black lines) for our early Mars scenario. Note the strong absorption bands of the atmospheric gases.}
    \label{fig: EarlyMarsSpec}
\end{figure}

Various studies have analyzed the potential climate of early Mars \citep[e.g.,][]{wordsworth2016, guzewich2021}. However, while some studies have modeled modern Mars-as-an-exoplanet \citep[e.g.,][]{wolfe2024}, less attention has been given to early-Mars-as-an-exoplanet. Fig.~\ref{fig: EarlyMarsSpec} displays $\pi{F_n}(\lambda)$ (top) and $P(\lambda)$ (bottom) for our early Mars model scenario (see Sect.~\ref{sec: EarlyMars} and Fig.~\ref{fig: Marsviews} for our maps) at $\alpha = 40\degree$ (solid red lines) and $90\degree$ (dashed black lines). Strong absorption bands of atmospheric CO$_2$, CH$_4$, and H$_2$O are seen in both $\pi{F_{n}}$ and $P$, caused by the large VMRs of the first two gases and the high absorption properties of H$_2$O, despite it being a trace gas in our simulation. At $\alpha = 90\degree$, Rayleigh scattering causes high levels of $P$ at shorter $\lambda$, while at $\alpha = 40\degree$, the larger surface area of the low-albedo ocean leads to a drop in $\pi{F_n}$ and subsequent rise in $P$ at longer $\lambda$.

\begin{figure}[ht!]
    \centering
    \includegraphics[width=\linewidth]{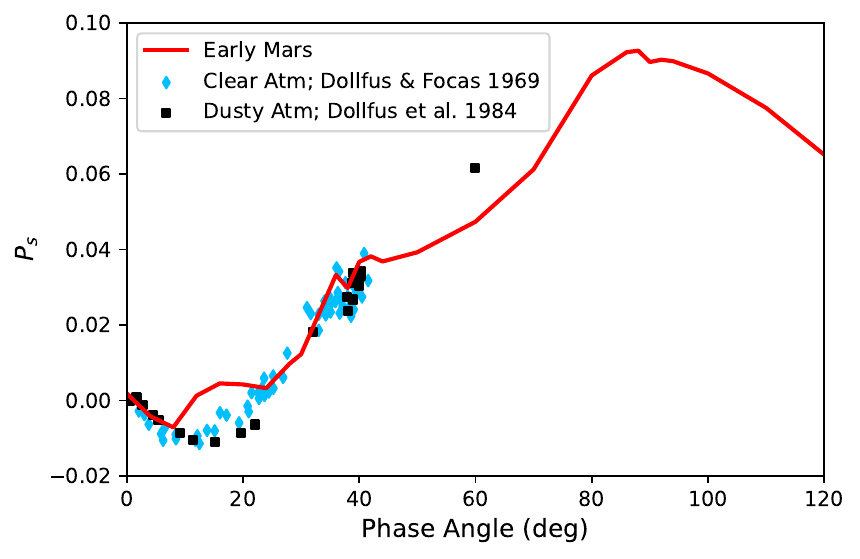}
    \caption{$P_s(\alpha)$ for our early Mars model at $\lambda = 0.5$ $\mu$m compared to archival observations of the polarization of modern Mars with a clear (at $\lambda = 0.51$ $\mu$m; blue diamonds) or dusty (at $\lambda = 0.52$ $\mu$m; black squares) atmosphere. Between $\alpha = 20\degree - 25\degree$, the polarization switches from being parallel (for smaller $\alpha$) to perpendicular (for larger $\alpha$) to the scattering plane.}
    \label{fig: MarsEarlyVsMod}
\end{figure}

In Fig.~\ref{fig: MarsEarlyVsMod} we plot our early Mars scenario at $\lambda = 0.5$ $\mu$m (solid red line) versus archival observations of modern Mars at $\lambda = 0.51$ $\mu$m from \citet[][]{dollfus1969} (blue diamonds) and $\lambda = 0.52$ $\mu$m from \citet[][]{dollfus1984} (black squares). Because the archival studies only observed the polarization in one orientation (i.e., the Stokes Q), we show here the resulting \textit{signed} degree of linear polarization, $P_s = -Q / F$, which also includes the direction of the polarization: if $P_s > 0$ ($P_s < 0$), the light is polarized perpendicular (parallel) to the planetary scattering plane. The measurements from \citet[][]{dollfus1969} are of areas of modern Mars where the atmosphere was clear and of negligible polarimetric contribution \citep[see, e.g.,][for a discussion]{ebisawa1993}, while the observations of \citet[][]{dollfus1984} were taken during a global dust storm in 1973 that dissipated over the course of the observations. The measurements between phase angles 0$\degree$ and 25$\degree$ in this latter case correspond to the period of maximum activity of the second 1973 global dust storm, while the Martian atmosphere progressively recovered its clear conditions at the larger phase angles. The measurement at $\alpha = 60\degree$ corresponds to observations taken by the VPM photopolarimeters on board the Soviet MARS-5 spacecraft.

Both of the modern observations have a negative polarization branch at small $\alpha$, indicative of regolith-like surfaces covered by small particles \citep[see, e.g.,][]{bagnulo2024}. While some differences exist between the clear and dusty conditions (i.e., a wider negative $P_s$ branch for the dusty atmosphere), the general form of the curves remains the same in both conditions. This implies that $P_s(\alpha)$ is dominated by dust and so it makes little difference whether the dust is settled on the surface or lofted into the atmosphere during a storm. Our early Mars model shows similar $P_s(\alpha)$ to the modern observations but with two additional features: the ``glory" feature at $\alpha \approx 15\degree$ and the primary rainbow feature at $\alpha \approx 40\degree$. Both of these are due to the presence of the water clouds, which were much more abundant and optically thicker on early Mars than modern Mars. We acknowledge, however, that if our early Mars model incorporated larger and/or nonspherical water ice cloud particles, the glory feature would most likely be reduced or even blurred out entirely, as is the case for observations of terrestrial water ice clouds \citep[e.g.][]{macke1996, hess1998}. Additionally, our early Mars phase curve shows $\sim$2\% less $P$ at $\alpha = 60\degree$ than the modern Mars observations, most likely caused by the increased multiple scattering within the early Mars water clouds, leading to lower $P$ \citep[e.g.,][]{stam2008}, compared to the scattering off the irregularly shaped dust grains settling out of the modern atmosphere, leading to higher $P$ \citep[e.g.,][]{santer1985}. Our results indicate that an early Mars should be easily distinguishable from a modern Mars due to the absence of dust and increased abundance of global water cloud coverage. Our early Mars model signatures in both unpolarized and polarized light are available in our Zenodo database.

\section{Observational Constraints} \label{sec: Observe}

Determining which types of potentially habitable exoplanets will be detectable, and under what conditions, is vital to maximize observational time for future missions aimed at characterizing these worlds \citep[][]{Astro2020}. \citet[][]{vaughan2023} provided preliminary analyses of Inner Working Angles (IWAs) and planet-to-star flux contrast ratios in both unpolarized and polarized light required by HWO for the characterization of specific features of modern-day Earth-like exoplanets around stars of multiple stellar types. \citet[][]{goodisgordon2025} expanded on this study to include contrast ratios as a function of $\lambda$ for the full VNIR range for multiple phase angles and for Earth-like planets across multiple points in their geological evolution.

Here we expand upon these two previous studies by determining the detectability of our model planets for different types of heterogeneities, as well as for the potentially habitable case of early wet Mars. Following the framework of \citet[][]{goodisgordon2025}, we calculated the unpolarized contrast ratios, $F{_p} / F{_s}$, for a planet at an orbital distance $a$ away from its host star as

\begin{eqnarray}
\frac{F_p}{F_s} = \frac{1}{4} \textbf{S}(\lambda, \alpha) \left({\frac{r}{a}}\right)^2.
\label{eq: contrastratio}
\end{eqnarray}

The polarized contrast ratios, $F{_{pol}} / F{_s}$, are then obtained by multiplying the $F{_p} / F{_s}$ for each model by its corresponding $P$ (see Equation~\ref{eq:degofpol}). As in \citet[][]{goodisgordon2025}, we assume unpolarized incident starlight, a planetary radius $r$ = $R_{\oplus}$, an orbital distance $a$ = 1 au, circular edge-on orbits (i.e., $i = 90\degree$), and perfect removal of any starlight (including all noise) from the planetary pixel.

\begin{figure*}[ht!]
    \centering
    \includegraphics[width=14.5cm]{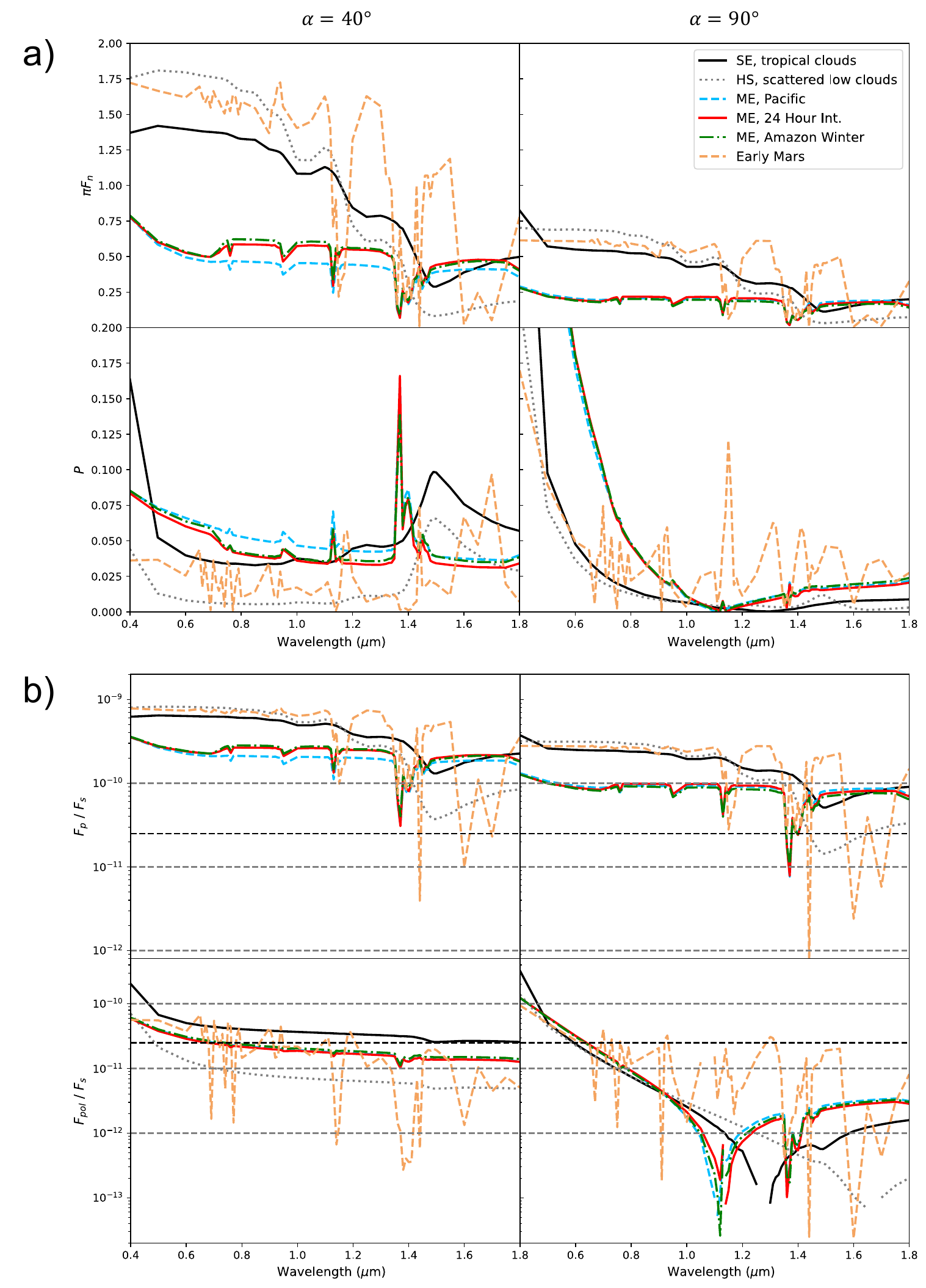}
    \caption{(a) $\pi{F_{n}}$($\lambda$) (top row) and $P(\lambda)$ (bottom row) for five cases of our early (SE: ``Slushball Earth" and HS: ``Hard Snowball") and modern Earth (ME) scenarios as well as our early Mars scenario, with their atmospheres and surfaces containing various heterogeneities. The spectra are plotted at either $\alpha = 40\degree$ (left column) or $\alpha = 90\degree$ (right column). (b) Unpolarized (top row) and polarized (bottom row) contrast ratios for the six models of panel (a). These contrast ratios assume the planets are in circular edge-on orbits at 1 au from the star. Also marked are contrast ratio limits of $1 \times 10^{-10}$, $1 \times 10^{-11}$, and $1 \times 10^{-12}$ (dashed gray lines), as well as the preliminary HWO lower limit of $2.5 \times 10^{-11}$ (dashed black lines). The breaks in some of the $F_{pol} / F{_s}$ curves at $\alpha = 90\degree$ are due to the polarized light changing directions.}
    \label{fig: ContrastRatios}
\end{figure*}

Panel (a) of Fig.~\ref{fig: ContrastRatios} shows $\pi{F_{n}}(\lambda)$ (top row) and $P(\lambda)$ (bottom row) of various cases of our Snowball and modern Earth scenarios as well as our early Mars scenario, while panel (b) shows the resulting unpolarized (top row) and polarized (bottom row) contrast ratios for these cases. The spectra are displayed at two key phase angles: $\alpha = 40\degree$ to capture the planets at the peak of the water cloud rainbow feature (left columns), and $\alpha = 90\degree$ to capture the planets at their widest separation from their host stars (left columns). Panel (b) also includes contrast ratio limits ranging from 1 $\times$ 10$^{-10}$ to 1 $\times$ 10$^{-12}$ (dashed gray lines) as well as the preliminary HWO lower contrast ratio limit of 2.5 $\times$ 10$^{-11}$ \citep[][black dashed lines]{mamajek2024}\footnote{Available online: \url{https://exoplanetarchive.ipac.caltech.edu/docs/2645_NASA_ExEP_Target_List_HWO_Documentation_2023.pdf}}.

The larger visible and illuminated disk at $\alpha = 40\degree$ results in larger $F{_p} / F{_s}$ for all six model cases shown here. At this phase, every model is detectable above the HWO lower contrast limit of 2.5 $\times$ 10$^{-11}$ across the full spectrum, with the exception of the deepest NIR absorption bands in the early Mars scenario. The VRE of the modern Earth (ME) Amazon winter view and the full 24-hour integration ME model are easily observed at this phase, and these two cases are easily separable from the ME Pacific view which displays no VRE. The early Earth and Mars models show similar contrasts in the VIS wavelengths but lower contrasts with larger separation and variation in the NIR wavelengths.

At $\alpha$ = 90$\degree$, the three ME cases become nearly indistinguishable but are still observable above the HWO contrast limit, except for the full depth of the 1.35 $\mu$m H$_2$O feature. Similarly, the cloudy SE case has $F{_p} / F{_s}$ above the HWO contrast limit. However, the deep NIR CO$_2$ and CH$_4$ absorption bands of the early Mars model as well as the NIR H$_2$O band of the cloudy HS case require contrasts lower than the HWO limit to resolve. Our results suggest that observations at both $\alpha$'s would allow for the detection of our model planets, with $\alpha = 40\degree$ allowing for their optimal characterization. We do acknowledge that at this phase angle the planets would be closer to their host star and would be more difficult to observe, requiring smaller IWAs. However, pushing for these smaller IWAs would allow us to resolve and distinguish the different cases shown here.

Similar to the contrast ratios presented in \citet[][]{vaughan2023} and \citet[][]{goodisgordon2025}, the resulting $F{_{pol}} / F{_s}$ for all six cases are lower than their corresponding unpolarized contrasts for both $\alpha$'s, with only the cloudy SE case being detectable above the HWO lower contrast limit across the full spectrum and only at $\alpha = 40\degree$. While the $F{_{pol}} / F{_s}$ for all six cases at this phase have higher contrasts than those at $\alpha = 90\degree$, the models are nearly indistinguishable, with all three ME cases overlapping with the continuum of the early Mars case and only the cloudy HS case showing any significant separation. On the other hand, at $\alpha = 90\degree$ the $F{_{pol}} / F{_s}$ show more defining features for all six cases, which allow the planetary scenarios to be differentiated. 

In particular, the breaks in the ME cases and early Mars case between 1.1 - 1.2 $\mu$m, the cloudy SE case at $\sim$1.3 $\mu$m, and the cloudy HS case at $\sim$1.65 $\mu$m are not due to atmospheric absorption but rather to changes in the direction of the polarization, with the light switching from being polarized perpendicular (for shorter $\lambda$s) to parallel (for longer $\lambda$s) to the scattering plane. These changes in the polarization direction are due to the presence of the clouds in the atmospheres of the different planetary scenarios. Our results indicate that the heterogeneity in the locations, distributions, and optical properties of the clouds in our model planets lead to varying wavelengths at which the direction of polarization changes for our models, therefore allowing us to use these contrast ratio features as a proxy for determining whether an observed planet could possess an environment more akin to modern Earth (with a direction change around 1.1 - 1.2 $\mu$m) or more extreme (with a direction change at longer $\lambda$).

\section{Discussion And Conclusions} \label{sec: DiscussConclude}

While most models of terrestrial exoplanets to date treat the planets as homogeneous objects for simplicity, in reality, Earth-like planets are dynamic systems with time-variable atmospheric and surface properties. These changing features can have noticeable effects on their resulting signatures, especially in polarized light where the vector nature of polarimetry makes it extremely sensitive to the locations of various features. Properly taking into account planetary heterogeneities in theoretical models of exoplanets will be vital for comparing models to near-future observations and improving the characterizations of potentially habitable worlds. Here, we presented the spectropolarimetric signatures of early and modern Earth with various types of planetary heterogeneities, as well as, to our knowledge, the first unpolarized and polarized signals of a heterogeneous early wet Mars. All models presented in this study are publicly available through Zenodo: \dataset[doi:10.5281/zenodo.17039224]{\doi{10.5281/zenodo.17039224}}.

The distributions of different surfaces and their locations on the observed disk of the planet play a larger role on the degree of polarization, $P$, than on the reflected flux, $\pi{F_{n}}$, for the models studied here.
For example, in Fig.~\ref{fig: ModEarthspecs} the $\pi{F_{n}}$ of the Amazon and Africa views of our modern Earth show very similar heights in their respective VRE signals. However, since the Amazon view has more vegetated surfaces located near the center of the disk, while the Africa view has more vegetated surfaces near the limbs, the vegetation in the Amazon view contributes more to the disk-integrated polarized signals. Therefore, the resulting VRE is larger in $P$ for the former versus the latter \textbf{case}. In a similar manner, the introduction of a tropical ocean belt (i.e., the ``Slushball Earth" case) on our Snowball Earth scenario breaks the homogeneity of the planet and causes additional variations in both $\pi{F_{n}}$ and $P$ for these models. Clear differences between this ``Slushball Earth" case and our ``Hard Snowball" case can be seen in the phase curves of Fig.~\ref{fig: SnowSlushphases}, where the signals in $P$ show much more variability and separation than those in $\pi{F_{n}}$. Our results highlight the complimentary role of polarimetric observations to traditional flux observations, and show the importance of acquiring both types of observations for the better characterization of exoplanets.

While the surface distributions of our models are heterogeneous and time-variable, we acknowledge that they are still simplified maps that only use six distinct wavelength-dependent albedos for the planetary pixels. As discussed in \citet[][]{roccetti2025}, this approach can lead to biases in our surface reflectances as we do not account for, e.g., multiple types of plants in our forest and grass surface pixels or dust-covered snow in our snow and ice surface pixels. To address this issue, \citet[][]{roccetti2024} developed the HAMSTER hyperspectral albedo map dataset, based on years of MODIS observations, which better captures the variability of Earth’s surface albedos. \citet[][]{roccetti2024} found that representing vegetated surfaces as a mixture of soils and plants leads to a substantially reduced VRE feature that has been overestimated in previous Earth-like simulations. 
However, these HAMSTER maps are specific to the modern Earth. Different surface mixtures (e.g., different plant types) might have existed at different times throughout the Earth's evolution or on an exoplanet. To keep our models generic and to follow previous studies \citep[e.g.,][]{gordon2023, goodisgordon2025}, we opted to use the single surface albedos from the NASA JPL EcoStress library instead of the HAMSTER database. This also allowed for faster computational runtimes for our models as more than one pixel could have similar surface and atmospheric properties. Additionally, as \citet[][]{roccetti2025} showed, the introduction of clouds and cloud inter-pixel and temporal variability in the models has a more significant effect on the disk-integrated signal than the surface albedo alone. Therefore, our choice of surface albedo should have mainly affected our clear atmosphere models and not the cloudy ones, and the large cloud coverage in our simulations should have stronger influences on our resulting disk-integrated signals.

While all of our model surfaces are treated as Lambertian, in reality surfaces do not reflect light isotropically. Rather, surface reflectance depends on illumination and viewing geometry, which is better captured using bidirectional reflection (or polarization) distribution functions \citep[BRDF/BPDF; e.g.,][]{maignan2009, breon2017}. For example, the rough and wavy nature of ocean surfaces can be modeled more accurately with approximations that take into account Fresnel reflections \citep[e.g.,][]{cox1954} and therefore better simulate the glint feature \citep[e.g.,][]{treesstam2019}. However, since the disk-integrated signals of exoplanets are expected to smear the surface reflectance, Lambertian surfaces provide sufficient approximations \citep[see, e.g.,][]{stam2008, berzosamolina2018, kopparla2018, gordon2023}. Additionally, since our models focus mostly on spectral characterizations between shorter and intermediate phase angles (i.e., $30\degree \leq \alpha \leq 120\degree$), our models are not as affected by the ocean glint, which primarily affects signals observed at longer phase angles \citep[$\alpha \approx 130\degree - 160\degree$, e.g.,][]{vaughan2023}. For example, \citet[][]{treesstam2019} showed that the signatures of a Fresnel-reflecting ocean planet are similar to those of a planet with a zero albedo (i.e., black) ocean up to $\alpha \approx 120\degree$, especially with increasing patchy cloud coverage that can sometimes block out the glint signal. Therefore, using BRDF/BPDF to better capture effects such as the ocean glint was not needed for our models and we opted to use Lambertian surfaces.

The diurnal rotation of Earth has been shown to have clear effects on the resulting $\pi{F_{n}}$ from the planet \citep[e.g.,][]{livengood2011}. Here we show that the same holds true for the disk-integrated $P$ for our Earth-like planets. More importantly, our models show larger separations between the diurnal signals at various wavelengths in $P$ than in $\pi{F_{n}}$, for both a simplified clear atmosphere planet (Fig.~\ref{fig: SCmodels}) and a realistic heterogeneous modern Earth (Fig.~\ref{fig: ModEarthrotate}). Additionally, the wavelength independence of the variability in the polarized diurnal light curves indicates that broadband observations at only a single bandpass would be needed to acquire an understanding of the planetary rotation. 

The diurnal variability introduces ``noise" into both the unpolarized and polarized phase curves of our modern Earth scenario when observed at short time steps, especially for $\pi{F_{n}}$ at longer $\lambda$ (see Fig.~\ref{fig: ModEarthInt}). Observing and characterizing this ``noise" will be important for future mapping efforts and the characterization of the long-term habitability of exoplanets. However, while the variable diurnal features disappear with longer integrations, the polarized phase curves still retain information about the atmosphere and water clouds of the Earth, even with integration times as high as $\sim$96 hours (not shown here). This retention of features would therefore allow us to place constraints on the planetary habitability even with longer exposures. \citet[][]{wolfe2024}, e.g., calculated that observing and characterizing terrestrial exoplanets in reflected flux at $\alpha = 90\degree$ ($\alpha = 45\degree$) with a signal-to-noise ratio of 5 requires exposure times of at least 250 (50) hours for planets located 10 pc away. They showed that the inclusion of planetary heterogeneities in their models, such as atmospheric dust, can shorten these exposure times. Because only a fraction of the reflected light is polarized, we would expect these integration times to be longer when observing in polarized light. HWO is planned to adopt the same exposure time limit as the LUVOIR \citep[][]{luvoir2019} and HabEx \citep[][]{gaudi2020} studies of 60 days \citep[][]{mamajek2024}. However, integration times that are too long can lead to errors in spectral characterization as well observing schedule conflicts \citep[e.g.,][]{stark2024}. Therefore, while our models show that longer integration times can still retain important habitability features of the planet in polarized light, future studies should aim to determine the shortest exposure times necessary by HWO to make these characterizations possible.

Our results for the seasonal effects of weather and surface patterns showed the ability of polarization to distinguish different planetary cases in ways that reflected flux alone cannot. While the different seasonal distributions of clouds on our ``Hard Snowball" Earth led to negligible changes in the resulting $\pi{F_{n}}$ (see Fig.~\ref{fig: SnowWinSumSpec} and \ref{fig: SnowWinSumPhase}), clear differences appear among the signals in $P$. At the same time, our models of the diurnal rotation of clouds over the Amazon (Fig.~\ref{fig: ModAmazAtmCircSpec} and \ref{fig: ModAmazAtmCircPhase}) highlight how polarization is less sensitive than flux to smaller-scale (i.e., daily) weather patterns, meaning less information will be lost in polarized observations of these planets when integrated over longer periods of time.

Apart from our early and modern Earth models, we modeled the unpolarized and polarized signatures of an early wet (and thus potentially habitable) Mars. We found that the warmer atmospheric and surface temperatures, as well as larger atmospheric gas VMRs, led to stronger absorption features in our early Mars spectra than those typically found in the spectra of modern Mars (see, e.g., the reflected flux and polarization signatures modeled in \citet[][]{stam2008b}). Additionally, the changing surface coverages of soil vs. open ocean in the illuminated disk between the $\alpha = 40\degree$ and $\alpha = 90\degree$ views led to noticeable differences in the resulting $P(\lambda)$. Comparing the polarized phase curve of our early Mars model with modern Mars observations of varying atmospheric conditions (i.e., both clear and dusty) across multiple $\alpha$, we found that the increased coverage of water clouds and lack of dust (either in the atmosphere or on the surface) in our early Mars simulation led to clear differences in the resulting disk-integrated polarization, especially for $\alpha < 25\degree$ (see Fig.~\ref{fig: MarsEarlyVsMod}). However, as these phase angles lie within the expected lower IWA for HWO \citep[e.g.,][]{vaughan2023, wolfe2024}, these differences might not be observable with currently planned instrumentation. On the other hand, the water cloud rainbow feature around $\alpha \approx 40\degree$ in our early Mars model, as well as the $\sim$2\% lower $P$ compared to the modern observation at $\alpha \approx 60\degree$ due to lack of dust, should be observable. Our results therefore indicate that an early Mars-like planet should still be distinguishable from a modern Mars-like planet in polarization with HWO.

Throughout this study, we have demonstrated how polarization provides a more robust method of planetary characterization than reflected flux alone. Although some defining features of all six planetary cases described in Sect.~\ref{sec: Observe} can be observed in both unpolarized and polarized light, only the $F{_p} / F{_s}$ contrasts can be characterized above the preliminary HWO lower contrast limit of 2.5 $\times$ 10$^{-11}$ (see Fig.~\ref{fig: ContrastRatios}). However, most of these contrasts, for both observed $\alpha$, show little separation between the unpolarized signals, especially at $\alpha = 90\degree$ for which we expect most future observations to be taken, since this allows for the widest separation between the planet and its host star. On the other hand, the $F{_{pol}} / F{_s}$ at $\alpha = 90\degree$ show significant differences among all of the tested cases. While our calculated contrast ratios did not take into account zodiacal or exozodiacal light, we did assume a background noise based on the JWST background model at a reference wavelength of 0.64 $\mu$m \citep[][]{rigby2023}. However, zodiacal and exozodiacal light can lead to large sources of noise in directly observed terrestrial exoplanets \citep[e.g.,][]{roberge2012, currie2023} and can potentially lower the observed contrast ratios. Taking into account these additional noise sources when calculating both unpolarized and polarized observing constraints is part of ongoing work.

Even though polarimetry requires achieving lower contrast ratios for our models, the diagnostic ability of the polarization to distinguish the various planets necessitates achieving these smaller contrasts. Pushing the HWO contrast limit floor down to 1 $\times$ 10$^{-13}$, similar to the lower contrast determined by the models of \citet[][]{goodisgordon2025}, would allow the community to resolve various heterogeneous features on multiple types of Earth-like planets, thus enabling us to better distinguish habitable vs. non-habitable planets and improve the future characterizations of these worlds.

\begin{acknowledgments}
We would like to thank Dr. Scott Guzewich for his assistance with our early Mars model inputs as well as Dr. Jeremy Bailey for his advice on our modeling efforts. K.G.G., T.K., K.B., C.V., and M.M-B. acknowledge the support of NASA Habitable Worlds grant No. 80NSSC24K0074. K.G.G. and T.K. acknowledge the University of Central Florida Advanced Research Computing Center high-performance computing resources made available for conducting the research reported in this paper (\url{https://arcc.ist.ucf.edu}). The results reported herein benefited from collaborations and/or information exchange within NASA's Nexus for Exoplanet System Science (NExSS) research coordination network sponsored by NASA's Science Mission Directorate. Finally, we thank the anonymous referees for helpful comments and suggestions, which resulted in improvements to our manuscript.
\end{acknowledgments}

\bibliography{Hetero}{}

\begin{thebibliography}{}
\expandafter\ifx\csname natexlab\endcsname\relax\def\natexlab#1{#1}\fi
\providecommand{\url}[1]{\href{#1}{#1}}
\providecommand{\dodoi}[1]{doi:~\href{http://doi.org/#1}{\nolinkurl{#1}}}
\providecommand{\doeprint}[1]{\href{http://ascl.net/#1}{\nolinkurl{http://ascl.net/#1}}}
\providecommand{\doarXiv}[1]{\href{https://arxiv.org/abs/#1}{\nolinkurl{https://arxiv.org/abs/#1}}}

\bibitem[{Abbot(2014)}]{abbot2014}
Abbot, D.~S. 2014, Journal of Climate, 27, 4391

\bibitem[{Abbot {et~al.}(2012)Abbot, Voigt, Branson, Pierrehumbert, Pollard, Le~Hir, \& Koll}]{abbot2012}
Abbot, D.~S., Voigt, A., Branson, M., {et~al.} 2012, Geophysical Research Letters, 39

\bibitem[{Abbot {et~al.}(2011)Abbot, Voigt, \& Koll}]{abbot2011}
Abbot, D.~S., Voigt, A., \& Koll, D. 2011, Journal of Geophysical Research: Atmospheres, 116

\bibitem[{Arney {et~al.}(2018)Arney, Domagal-Goldman, \& Meadows}]{arney2018}
Arney, G., Domagal-Goldman, S.~D., \& Meadows, V.~S. 2018, Astrobiology, 18, 311

\bibitem[{Arney {et~al.}(2016)Arney, Domagal-Goldman, Meadows, Wolf, Schwieterman, Charnay, Claire, H{\'e}brard, \& Trainer}]{arney2016}
Arney, G., Domagal-Goldman, S.~D., Meadows, V.~S., {et~al.} 2016, Astrobiology, 16, 873

\bibitem[{Arney {et~al.}(2017)Arney, Meadows, Domagal-Goldman, Deming, Robinson, Tovar, Wolf, \& Schwieterman}]{arney2017}
Arney, G.~N., Meadows, V.~S., Domagal-Goldman, S.~D., {et~al.} 2017, The Astrophysical Journal, 836, 49

\bibitem[{Bagnulo {et~al.}(2024)Bagnulo, Belskaya, Cellino, Kwon, Mu{\~n}oz, \& Stam}]{bagnulo2024}
Bagnulo, S., Belskaya, I., Cellino, A., {et~al.} 2024, The Astronomy and Astrophysics Review, 32, 1

\bibitem[{Bailey(2007)}]{bailey2007}
Bailey, J. 2007, Astrobiology, 7, 320

\bibitem[{Baldridge {et~al.}(2009)Baldridge, Hook, Grove, \& Rivera}]{baldridge2009}
Baldridge, A.~M., Hook, S.~J., Grove, C.~I., \& Rivera, G. 2009, Remote Sensing of Environment, 113, 711

\bibitem[{Bar-Nun {et~al.}(1988)Bar-Nun, Kleinfeld, \& Ganor}]{bar1988}
Bar-Nun, A., Kleinfeld, I., \& Ganor, E. 1988, Journal of Geophysical Research: Atmospheres, 93, 8383

\bibitem[{Berzosa~Molina {et~al.}(2018)Berzosa~Molina, Rossi, \& Stam}]{berzosamolina2018}
Berzosa~Molina, J., Rossi, L., \& Stam, D.~M. 2018, Astronomy \& Astrophysics, 618, A162

\bibitem[{Bouley {et~al.}(2016)Bouley, Baratoux, Matsuyama, Forget, S{\'e}journ{\'e}, Turbet, \& Costard}]{bouley2016}
Bouley, S., Baratoux, D., Matsuyama, I., {et~al.} 2016, Nature, 531, 344

\bibitem[{Braun {et~al.}(2022)Braun, H{\"o}rner, Voigt, \& Pinto}]{braun2022}
Braun, C., H{\"o}rner, J., Voigt, A., \& Pinto, J.~G. 2022, Nature Geoscience, 15, 489

\bibitem[{Breon \& Maignan(2017)}]{breon2017}
Breon, F.-M., \& Maignan, F. 2017, Earth System Science Data, 9, 31

\bibitem[{Byrne(2009)}]{byrne2009}
Byrne, S. 2009, Annual Review of Earth and Planetary Sciences, 37, 535

\bibitem[{Cantor {et~al.}(2001)Cantor, James, Caplinger, \& Wolff}]{cantor2001}
Cantor, B.~A., James, P.~B., Caplinger, M., \& Wolff, M.~J. 2001, Journal of Geophysical Research: Planets, 106, 23653

\bibitem[{Catling \& Zahnle(2020)}]{catling2020}
Catling, D.~C., \& Zahnle, K.~J. 2020, Science advances, 6, eaax1420

\bibitem[{Chubb {et~al.}(2024)Chubb, Stam, Helling, Samra, \& Carone}]{chubb2024}
Chubb, K.~L., Stam, D.~M., Helling, C., Samra, D., \& Carone, L. 2024, Monthly Notices of the Royal Astronomical Society, 527, 4955

\bibitem[{Claire {et~al.}(2012)Claire, Sheets, Cohen, Ribas, Meadows, \& Catling}]{claire2012}
Claire, M.~W., Sheets, J., Cohen, M., {et~al.} 2012, The Astrophysical Journal, 757, 95

\bibitem[{Cotton {et~al.}(2017)Cotton, Marshall, Bailey, Kedziora-Chudczer, Bott, Marsden, \& Carter}]{cotton2017}
Cotton, D.~V., Marshall, J.~P., Bailey, J., {et~al.} 2017, Monthly Notices of the Royal Astronomical Society, 467, 873

\bibitem[{Cowan {et~al.}(2012)Cowan, Abbot, \& Voigt}]{cowan2012}
Cowan, N.~B., Abbot, D.~S., \& Voigt, A. 2012, The Astrophysical Journal Letters, 752, L3

\bibitem[{Cowan {et~al.}(2009)Cowan, Agol, Meadows, Robinson, Livengood, Deming, Lisse, A'Hearn, Wellnitz, Seager, {et~al.}}]{cowan2009}
Cowan, N.~B., Agol, E., Meadows, V.~S., {et~al.} 2009, The Astrophysical Journal, 700, 915

\bibitem[{{Cowan} {et~al.}(2011){Cowan}, {Robinson}, {Livengood}, {Deming}, {Agol}, {A'Hearn}, {Charbonneau}, {Lisse}, {Meadows}, {Seager}, {Shields}, \& {Wellnitz}}]{cowan2011}
{Cowan}, N.~B., {Robinson}, T., {Livengood}, T.~A., {et~al.} 2011, \apj, 731, 76, \dodoi{10.1088/0004-637X/731/1/76}

\bibitem[{Cox \& Munk(1954)}]{cox1954}
Cox, C., \& Munk, W. 1954, Journal of the Optical Society of America, 44, 838

\bibitem[{Currie {et~al.}(2023)Currie, Stark, Kammerer, Juanola-Parramon, \& Meadows}]{currie2023}
Currie, M.~H., Stark, C.~C., Kammerer, J., Juanola-Parramon, R., \& Meadows, V.~S. 2023, The Astronomical Journal, 166, 197

\bibitem[{De~Kok {et~al.}(2011)De~Kok, Stam, \& Karalidi}]{dekok2011}
De~Kok, R.~J., Stam, D.~M., \& Karalidi, T. 2011, The Astrophysical Journal, 741, 59

\bibitem[{De~Rooij \& Van~der Stap(1984)}]{derooijvanderstap1984}
De~Rooij, W.~A., \& Van~der Stap, C. C. A.~H. 1984, Astronomy and Astrophysics, 131, 237

\bibitem[{Dollfus {et~al.}(1984)Dollfus, Bowell, \& Ebisawa}]{dollfus1984}
Dollfus, A., Bowell, E., \& Ebisawa, S. 1984, Astronomy and Astrophysics (ISSN 0004-6361), 134 no. 2, 343

\bibitem[{Dollfus \& Focas(1969)}]{dollfus1969}
Dollfus, A., \& Focas, J. 1969, Astronomy and Astrophysics, 2, 63

\bibitem[{Ebisawa \& Dollfus(1993)}]{ebisawa1993}
Ebisawa, S., \& Dollfus, A. 1993, Astronomy and Astrophysics, 272, 671

\bibitem[{Emde \& Mayer(2018)}]{emde2018}
Emde, C., \& Mayer, B. 2018, Journal of Quantitative Spectroscopy and Radiative Transfer, 218, 151

\bibitem[{Foley(2015)}]{foley2015}
Foley, B.~J. 2015, The Astrophysical Journal, 812, 36

\bibitem[{Gaudi {et~al.}(2020)Gaudi, Seager, Mennesson, Kiessling, Warfield, Cahoy, Clarke, Domagal-Goldman, Feinberg, Guyon, {et~al.}}]{gaudi2020}
Gaudi, B.~S., Seager, S., Mennesson, B., {et~al.} 2020, arXiv preprint arXiv:2001.06683

\bibitem[{{Ge} {et~al.}(2019){Ge}, {Zhang}, {Fletcher}, {Orton}, {Sinclair}, {Fernandes}, {Momary}, {Kasaba}, {Sato}, \& {Fujiyoshi}}]{ge2019}
{Ge}, H., {Zhang}, X., {Fletcher}, L.~N., {et~al.} 2019, AJ, 157, 89, \dodoi{10.3847/1538-3881/aafba7}

\bibitem[{Glaser {et~al.}(2020)Glaser, Hartnett, Desch, Unterborn, Anbar, Buessecker, Fisher, Glaser, Kane, Lisse, {et~al.}}]{glaser2020}
Glaser, D.~M., Hartnett, H.~E., Desch, S.~J., {et~al.} 2020, The Astrophysical Journal, 893, 163

\bibitem[{{Gomez Barrientos} {et~al.}(2023){Gomez Barrientos}, {MacDonald}, {Lewis}, \& {Kaltenegger}}]{gomezbarrientos2023}
{Gomez Barrientos}, J., {MacDonald}, R.~J., {Lewis}, N.~K., \& {Kaltenegger}, L. 2023, arXiv e-prints, arXiv:2301.01775.
\newblock \doarXiv{2301.01775}

\bibitem[{Goodis~Gordon {et~al.}(2025)Goodis~Gordon, Karalidi, Bott, Wogan, Arney, Parenteau, Kataria, \& Meadows}]{goodisgordon2025}
Goodis~Gordon, K.~E., Karalidi, T., Bott, K.~M., {et~al.} 2025, The Astrophysical Journal, 983, 168

\bibitem[{Gordon {et~al.}(2023)Gordon, Karalidi, Bott, Miles-P{\'a}ez, Mulder, \& Stam}]{gordon2023}
Gordon, K.~E., Karalidi, T., Bott, K.~M., {et~al.} 2023, The Astrophysical Journal, 945, 166

\bibitem[{Guzewich \& Smith(2019)}]{guzewich2019}
Guzewich, S.~D., \& Smith, M.~D. 2019, Journal of Geophysical Research: Planets, 124, 636

\bibitem[{Guzewich {et~al.}(2014)Guzewich, Smith, \& Wolff}]{guzewich2014}
Guzewich, S.~D., Smith, M.~D., \& Wolff, M.~J. 2014, Journal of Geophysical Research: Planets, 119, 2694

\bibitem[{Guzewich {et~al.}(2021)Guzewich, Way, Aleinov, Wolf, Del~Genio, Wordsworth, \& Tsigaridis}]{guzewich2021}
Guzewich, S.~D., Way, M.~J., Aleinov, I., {et~al.} 2021, Journal of Geophysical Research: Planets, 126, e2021JE006825

\bibitem[{Hale \& Querry(1973)}]{hale1973}
Hale, G.~M., \& Querry, M.~R. 1973, Applied optics, 12, 555

\bibitem[{Hansen \& Hovenier(1974)}]{hansenhovenier1974}
Hansen, J.~E., \& Hovenier, J.~W. 1974, Journal of Atmospheric Sciences, 31, 1137

\bibitem[{Hansen \& Travis(1974)}]{hansentravis1974}
Hansen, J.~E., \& Travis, L.~D. 1974, Space science reviews, 16, 527

\bibitem[{Hasler {et~al.}(1976)Hasler, Shenk, \& Skillman}]{hasler1976}
Hasler, A.~F., Shenk, W., \& Skillman, W. 1976, Journal of Applied Meteorology and Climatology, 15, 10

\bibitem[{Hess {et~al.}(1998)Hess, Koelemeijer, \& Stammes}]{hess1998}
Hess, M., Koelemeijer, R. B.~A., \& Stammes, P. 1998, Journal of Quantitative Spectroscopy and Radiative Transfer, 60, 301

\bibitem[{Heymsfield \& Platt(1984)}]{heymsfield1984}
Heymsfield, A.~J., \& Platt, C. M.~R. 1984, Journal of Atmospheric Sciences, 41, 846

\bibitem[{Hoffman {et~al.}(1998)Hoffman, Kaufman, Halverson, \& Schrag}]{hoffman1998}
Hoffman, P.~F., Kaufman, A.~J., Halverson, G.~P., \& Schrag, D.~P. 1998, science, 281, 1342

\bibitem[{Hoffman {et~al.}(2017)Hoffman, Abbot, Ashkenazy, Benn, Brocks, Cohen, Cox, Creveling, Donnadieu, Erwin, {et~al.}}]{hoffman2017}
Hoffman, P.~F., Abbot, D.~S., Ashkenazy, Y., {et~al.} 2017, Science Advances, 3, e1600983

\bibitem[{Hovenier {et~al.}(2004)Hovenier, Van~der Mee, \& Domke}]{hovenier2004}
Hovenier, J.~W., Van~der Mee, C. V.~M., \& Domke, H. 2004, Transfer of Polarized Light in Planetary Atmospheres: Basic Concepts and Practical Methods, Vol. 318 (Dordrecht: Kluwer Academic Publishers (Springer))

\bibitem[{Hyde {et~al.}(2000)Hyde, Crowley, Baum, \& Peltier}]{hyde2000}
Hyde, W.~T., Crowley, T.~J., Baum, S.~K., \& Peltier, W.~R. 2000, Nature, 405, 425

\bibitem[{Ingersoll {et~al.}(2004)Ingersoll, Dowling, Gierasch, Orton, Read, S{\'a}nchez-Lavega, Showman, Simon-Miller, \& Vasavada}]{ingersoll2004}
Ingersoll, A.~P., Dowling, T.~E., Gierasch, P.~J., {et~al.} 2004, Jupiter: The planet, satellites and magnetosphere, 105

\bibitem[{Karalidi \& Stam(2012)}]{karalidistam2012}
Karalidi, T., \& Stam, D.~M. 2012, Astronomy \& Astrophysics, 546, A56

\bibitem[{Karalidi {et~al.}(2013)Karalidi, Stam, \& Guirado}]{karalidi2013flux}
Karalidi, T., Stam, D.~M., \& Guirado, D. 2013, Astronomy \& Astrophysics, 555, A127

\bibitem[{Karalidi {et~al.}(2011)Karalidi, Stam, \& Hovenier}]{karalidi2011}
Karalidi, T., Stam, D.~M., \& Hovenier, J.~W. 2011, Astronomy \& Astrophysics, 530, A69

\bibitem[{Karalidi {et~al.}(2012)Karalidi, Stam, \& Hovenier}]{karalidi2012rainbow}
---. 2012, Astronomy \& Astrophysics, 548, A90

\bibitem[{Kelkar {et~al.}(2025)Kelkar, Saxena, Kopparapu, \& Monteiro}]{kelkar2025}
Kelkar, S., Saxena, P., Kopparapu, R., \& Monteiro, J. 2025, The Planetary Science Journal, 6, 87

\bibitem[{Kemp {et~al.}(1987)Kemp, Henson, Steiner, \& Powell}]{kemp1987}
Kemp, J.~C., Henson, G.~D., Steiner, C.~T., \& Powell, E.~R. 1987, Nature, 326, 270

\bibitem[{King {et~al.}(2004)King, Closs, Spangler, Greenstone, Wharton, \& Myers}]{king2004}
King, M.~D., Closs, J., Spangler, S., {et~al.} 2004, EOS Data Products Handbook, Volume 1 (Greenbelt: NASA Goddard Space Flight Center)

\bibitem[{King {et~al.}(2013)King, Platnick, Menzel, Ackerman, \& Hubanks}]{king2013}
King, M.~D., Platnick, S., Menzel, W.~P., Ackerman, S.~A., \& Hubanks, P.~A. 2013, IEEE transactions on geoscience and remote sensing, 51, 3826

\bibitem[{Kirschvink(1992)}]{kirschvink1992}
Kirschvink, J.~L. 1992, in The Proterozoic Biosphere: A MultidisciplinaryStudy, ed. J.~W. Schopf \& C.~Klein (New York: Cambridge University Press), 51--52

\bibitem[{Kofman {et~al.}(2024)Kofman, Villanueva, Fauchez, Mandell, Johnson, Payne, Latouf, \& Kelkar}]{kofman2024}
Kofman, V., Villanueva, G.~L., Fauchez, T.~J., {et~al.} 2024, The Planetary Science Journal, 5, 197

\bibitem[{Kokaly {et~al.}(2017)Kokaly, Clark, Swayze, Livo, Hoefen, Pearson, Wise, Benzel, Lowers, Driscoll, {et~al.}}]{kokaly2017}
Kokaly, R.~F., Clark, R.~N., Swayze, G.~A., {et~al.} 2017, United States Geological Survey (USGS): Reston, VA, USA, 61

\bibitem[{Kopparapu {et~al.}(2016)Kopparapu, Wolf, Haqq-Misra, Yang, Kasting, Meadows, Terrien, \& Mahadevan}]{kopparapu2016}
Kopparapu, R.~K., Wolf, E.~T., Haqq-Misra, J., {et~al.} 2016, The Astrophysical Journal, 819, 84

\bibitem[{Kopparapu {et~al.}(2013)Kopparapu, Ramirez, Kasting, Eymet, Robinson, Mahadevan, Terrien, Domagal-Goldman, Meadows, \& Deshpande}]{kopparapu2013}
Kopparapu, R.~K., Ramirez, R., Kasting, J.~F., {et~al.} 2013, The Astrophysical Journal, 765, 131

\bibitem[{Kopparla {et~al.}(2018)Kopparla, Natraj, Crisp, Bott, Swain, \& Yung}]{kopparla2018}
Kopparla, P., Natraj, V., Crisp, D., {et~al.} 2018, The Astronomical Journal, 156, 143

\bibitem[{Lingam \& Loeb(2019)}]{lingam2019}
Lingam, M., \& Loeb, A. 2019, The Astronomical Journal, 157, 25

\bibitem[{Livengood {et~al.}(2011)Livengood, Deming, A'hearn, Charbonneau, Hewagama, Lisse, McFadden, Meadows, Robinson, Seager, {et~al.}}]{livengood2011}
Livengood, T.~A., Deming, L.~D., A'hearn, M.~F., {et~al.} 2011, Astrobiology, 11, 907

\bibitem[{Lustig-Yaeger {et~al.}(2018)Lustig-Yaeger, Meadows, Mendoza, Schwieterman, Fujii, Luger, \& Robinson}]{lustigyaeger2018}
Lustig-Yaeger, J., Meadows, V.~S., Mendoza, G.~T., {et~al.} 2018, The Astronomical Journal, 156, 301

\bibitem[{Macke {et~al.}(1996)Macke, Mueller, \& Raschke}]{macke1996}
Macke, A., Mueller, J., \& Raschke, E. 1996, Journal of Atmospheric Sciences, 53, 2813

\bibitem[{Maignan {et~al.}(2009)Maignan, Br{\'e}on, F{\'e}d{\`e}le, \& Bouvier}]{maignan2009}
Maignan, F., Br{\'e}on, F.-M., F{\'e}d{\`e}le, E., \& Bouvier, M. 2009, Remote Sensing of Environment, 113, 2642

\bibitem[{Mamajek \& Stapelfeldt(2024)}]{mamajek2024}
Mamajek, E., \& Stapelfeldt, K. 2024, arXiv preprint arXiv:2402.12414

\bibitem[{McLean {et~al.}(2017)McLean, Stam, Bagnulo, Borisov, Devog{\`e}le, Cellino, Rivet, Bendjoya, Vernet, Paolini, {et~al.}}]{mclean2017}
McLean, W., Stam, D.~M., Bagnulo, S., {et~al.} 2017, Astronomy \& Astrophysics, 601, A142

\bibitem[{Meerdink {et~al.}(2019)Meerdink, Hook, Roberts, \& Abbott}]{meerdink2019}
Meerdink, S.~K., Hook, S.~J., Roberts, D.~A., \& Abbott, E.~A. 2019, Remote Sensing of Environment, 230, 111196

\bibitem[{Miles-P{\'a}ez {et~al.}(2014)Miles-P{\'a}ez, Pall{\'e}, \& Osorio}]{miles2014}
Miles-P{\'a}ez, P.~A., Pall{\'e}, E., \& Osorio, M. R.~Z. 2014, Astronomy \& Astrophysics, 562, L5

\bibitem[{Mishchenko {et~al.}(1994)Mishchenko, Lacis, \& Travis}]{mishchenko1994}
Mishchenko, M.~I., Lacis, A.~A., \& Travis, L.~D. 1994, Journal of Quantitative Spectroscopy and Radiative Transfer, 51, 491

\bibitem[{Mitchell \& Lora(2016)}]{mitchell2016}
Mitchell, J.~L., \& Lora, J.~M. 2016, Annual Review of Earth and Planetary Sciences, 44, 353

\bibitem[{{National Academies of Sciences{,} Engineering{,} and Medicine} {et~al.}(2021)}]{Astro2020}
{National Academies of Sciences{,} Engineering{,} and Medicine}, {et~al.} 2021, Pathways to Discovery in Astronomy and Astrophysics for the 2020s

\bibitem[{Rigby {et~al.}(2023)Rigby, Lightsey, Mar{\'\i}n, Bowers, Smith, Glasse, McElwain, Rieke, Chary, Liu, {et~al.}}]{rigby2023}
Rigby, J.~R., Lightsey, P.~A., Mar{\'\i}n, M.~G., {et~al.} 2023, Publications of the Astronomical Society of the Pacific, 135, 048002

\bibitem[{Roberge {et~al.}(2012)Roberge, Chen, Millan-Gabet, Weinberger, Hinz, Stapelfeldt, Absil, Kuchner, \& Bryden}]{roberge2012}
Roberge, A., Chen, C.~H., Millan-Gabet, R., {et~al.} 2012, Publications of the Astronomical Society of the Pacific, 124, 799

\bibitem[{Robinson \& Reinhard(2018)}]{robinson2018}
Robinson, T.~D., \& Reinhard, C.~T. 2018, Planetary Astrobiology, 379

\bibitem[{Robinson {et~al.}(2011)Robinson, Meadows, Crisp, Deming, A'hearn, Charbonneau, Livengood, Seager, Barry, Hearty, {et~al.}}]{robinson2011}
Robinson, T.~D., Meadows, V.~S., Crisp, D., {et~al.} 2011, Astrobiology, 11, 393

\bibitem[{{Roccetti} {et~al.}(2024){Roccetti}, {Bugliaro}, {G{\"o}dde}, {Emde}, {Hamann}, {Manev}, {Sterzik}, \& {Wehrum}}]{roccetti2024}
{Roccetti}, G., {Bugliaro}, L., {G{\"o}dde}, F., {et~al.} 2024, Atmospheric Measurement Techniques, 17, 6025, \dodoi{10.5194/amt-17-6025-2024}

\bibitem[{Roccetti {et~al.}(2025)Roccetti, Emde, Sterzik, Manev, Seidel, \& Bagnulo}]{roccetti2025}
Roccetti, G., Emde, C., Sterzik, M.~F., {et~al.} 2025, arXiv preprint arXiv:2504.02048

\bibitem[{Rossi \& Stam(2018)}]{rossi2018}
Rossi, L., \& Stam, D.~M. 2018, Astronomy \& Astrophysics, 616, A117

\bibitem[{Ruddiman(2001)}]{ruddiman2001}
Ruddiman, W.~F. 2001, Earth's Climate: past and future (Macmillan)

\bibitem[{Sanrom{\'a} {et~al.}(2013)Sanrom{\'a}, Pall{\'e}, Parenteau, Kiang, Guti{\'e}rrez-Navarro, L{\'o}pez, \& Monta{\~n}{\'e}s-Rodr{\'\i}guez}]{sanroma2013}
Sanrom{\'a}, E., Pall{\'e}, E., Parenteau, M.~N., {et~al.} 2013, The Astrophysical Journal, 780, 52

\bibitem[{Santer {et~al.}(1985)Santer, Deschamps, Ksanfomaliti, \& Dollfus}]{santer1985}
Santer, R., Deschamps, M., Ksanfomaliti, L.~V., \& Dollfus, A. 1985, Astronomy and Astrophysics (ISSN 0004-6361), 150, no. 2, 217

\bibitem[{Schmid {et~al.}(2011)Schmid, Joos, Buenzli, \& Gisler}]{schmid2011}
Schmid, H.~M., Joos, F., Buenzli, E., \& Gisler, D. 2011, Icarus, 212, 701

\bibitem[{Sengupta \& Marley(2010)}]{sengupta2010}
Sengupta, S., \& Marley, M.~S. 2010, The Astrophysical Journal Letters, 722, L142

\bibitem[{Shkuratov {et~al.}(2005)Shkuratov, Kreslavsky, Kaydash, Videen, Bell~III, Wolff, Hubbard, Noll, \& Lubenow}]{shkuratov2005}
Shkuratov, Y., Kreslavsky, M., Kaydash, V., {et~al.} 2005, Icarus, 176, 1

\bibitem[{Stam(2008)}]{stam2008}
Stam, D.~M. 2008, Astronomy \& Astrophysics, 482, 989

\bibitem[{Stam \& Hovenier(2005)}]{stam2005}
Stam, D.~M., \& Hovenier, J.~W. 2005, Astronomy \& Astrophysics, 444, 275

\bibitem[{{Stam} {et~al.}(2008){Stam}, {Laan}, {Snik}, {Karalidi}, {Keller}, {Ter Horst}, {Navarro}, {Aas}, {de Vries}, {Oomen}, \& {Hoogeveen}}]{stam2008b}
{Stam}, D.~M., {Laan}, E., {Snik}, F., {et~al.} 2008, in Third International Workshop on The Mars Atmosphere: Modeling and Observations, Vol. 1447 (Williamsburg, Virginia: LPI Contributions), 9078

\bibitem[{Stark {et~al.}(2024)Stark, Mennesson, Bryson, Ford, Robinson, Belikov, Bolcar, Feinberg, Guyon, Latouf, {et~al.}}]{stark2024}
Stark, C.~C., Mennesson, B., Bryson, S., {et~al.} 2024, Journal of Astronomical Telescopes, Instruments, and Systems, 10, 034006

\bibitem[{Sterzik {et~al.}(2020)Sterzik, Bagnulo, Emde, \& Manev}]{sterzik2020}
Sterzik, M.~F., Bagnulo, S., Emde, C., \& Manev, M. 2020, Astronomy \& Astrophysics, 639, A89

\bibitem[{Stolker {et~al.}(2017)Stolker, Min, Stam, Molli{\`e}re, Dominik, \& Waters}]{stolker2017}
Stolker, T., Min, M., Stam, D.~M., {et~al.} 2017, Astronomy \& Astrophysics, 607, A42

\bibitem[{{The LUVOIR Team} {et~al.}(2019)}]{luvoir2019}
{The LUVOIR Team}, {et~al.} 2019, arXiv preprint arXiv:1912.06219

\bibitem[{Tilstra {et~al.}(2021)Tilstra, Tuinder, Wang, \& Stammes}]{tilstra2021}
Tilstra, L.~G., Tuinder, O. N.~E., Wang, P., \& Stammes, P. 2021, Atmospheric Measurement Techniques, 14, 4219

\bibitem[{Trees \& Stam(2019)}]{treesstam2019}
Trees, V. J.~H., \& Stam, D.~M. 2019, Astronomy \& Astrophysics, 626, A129

\bibitem[{Trees \& Stam(2022)}]{trees2022}
---. 2022, Astronomy \& Astrophysics, 664, A172

\bibitem[{Van~Diedenhoven {et~al.}(2007)Van~Diedenhoven, Hasekamp, \& Landgraf}]{vandiedenhoven2007}
Van~Diedenhoven, B., Hasekamp, O.~P., \& Landgraf, J. 2007, Journal of Geophysical Research: Atmospheres, 112, D15

\bibitem[{Vaughan {et~al.}(2023)Vaughan, Gebhard, Bott, Casewell, Cowan, Doelman, Kenworthy, Mazoyer, Millar-Blanchaer, Trees, {et~al.}}]{vaughan2023}
Vaughan, S.~R., Gebhard, T.~D., Bott, K., {et~al.} 2023, Monthly Notices of the Royal Astronomical Society, 524, 5477

\bibitem[{Warren \& Brandt(2008)}]{warren2008}
Warren, S.~G., \& Brandt, R.~E. 2008, Journal of Geophysical Research: Atmospheres, 113

\bibitem[{Way {et~al.}(2017)Way, Aleinov, Amundsen, Chandler, Clune, Del~Genio, Fujii, Kelley, Kiang, Sohl, {et~al.}}]{way2017}
Way, M.~J., Aleinov, I., Amundsen, D.~S., {et~al.} 2017, The Astrophysical Journal Supplement Series, 231, 12

\bibitem[{West {et~al.}(2022)West, Dumont, Hu, Natraj, Breckinridge, \& Chen}]{west2022}
West, R.~A., Dumont, P., Hu, R., {et~al.} 2022, The Astrophysical Journal, 940, 183

\bibitem[{West \& Smith(1991)}]{west1991}
West, R.~A., \& Smith, P.~H. 1991, Icarus, 90, 330

\bibitem[{Wogan {et~al.}(2023)Wogan, Catling, Zahnle, \& Lupu}]{wogan2023}
Wogan, N.~F., Catling, D.~C., Zahnle, K.~J., \& Lupu, R. 2023, The Planetary Science Journal, 4, 169

\bibitem[{Wolf {et~al.}(2017)Wolf, Shields, Kopparapu, Haqq-Misra, \& Toon}]{wolf2017}
Wolf, E.~T., Shields, A.~L., Kopparapu, R.~K., Haqq-Misra, J., \& Toon, O.~B. 2017, The Astrophysical Journal, 837, 107

\bibitem[{{Wolfe} \& {Robinson}(2024)}]{wolfe2024}
{Wolfe}, C.~A., \& {Robinson}, T.~D. 2024, \planss, 250, 105944, \dodoi{10.1016/j.pss.2024.105944}

\bibitem[{Wordsworth {et~al.}(2013)Wordsworth, Forget, Millour, Head, Madeleine, \& Charnay}]{wordsworth2013}
Wordsworth, R., Forget, F., Millour, E., {et~al.} 2013, Icarus, 222, 1

\bibitem[{Wordsworth {et~al.}(2017)Wordsworth, Kalugina, Lokshtanov, Vigasin, Ehlmann, Head, Sanders, \& Wang}]{wordsworth2017}
Wordsworth, R., Kalugina, Y., Lokshtanov, S., {et~al.} 2017, Geophysical Research Letters, 44, 665

\bibitem[{Wordsworth(2016)}]{wordsworth2016}
Wordsworth, R.~D. 2016, Annual Review of Earth and Planetary Sciences, 44, 381

\bibitem[{Yan \& Yang(2024)}]{yan2024}
Yan, M., \& Yang, J. 2024, Journal of Geophysical Research: Atmospheres, 129, e2023JD040688

\end{thebibliography}
\bibliographystyle{aasjournal}



\end{document}